\documentclass[aps,reprint,superscriptaddress,amsmath,amssymb,pra,longbibliography]{revtex4-2}

\newcommand{\Jscience}{Science}

\newcommand{\Jpra}{Phys. Rev. A}

\newcommand{\Jepl}{Europhys. Lett.}

\newcommand{\crasphy}{C. R. Phys.}

\usepackage{amsmath}
\usepackage{amssymb}
\usepackage{graphicx}
\usepackage{float}
\usepackage[english]{babel} 

\usepackage[usenames,dvipsnames]{xcolor}
\usepackage[colorlinks=true,citecolor=Cerulean,linkcolor=RubineRed,urlcolor=Cerulean]{hyperref}

\newcommand{\ket}[1]{|{#1}\rangle}

\newcommand{\kB}{k_{\textrm{\tiny B}}}
\newcommand{\Er}{E_{\textrm{r}}}

\begin{document}

\title{Effective tight-binding models in optical moir\'e potentials}

\author{Dean Johnstone}
\affiliation{CPHT, CNRS, Ecole Polytechnique, IP Paris, F-91128 Palaiseau, France}

\author{Shanya Mishra}
\affiliation{CPHT, CNRS, Ecole Polytechnique, IP Paris, F-91128 Palaiseau, France}

\author{Zhaoxuan Zhu}
\affiliation{CPHT, CNRS, Ecole Polytechnique, IP Paris, F-91128 Palaiseau, France}

\author{Laurent Sanchez-Palencia}
\affiliation{CPHT, CNRS, Ecole Polytechnique, IP Paris, F-91128 Palaiseau, France}

\date{\today}

\begin{abstract}
A twist between two systems offers the possibility to drastically change the underlying physical properties. To that end, we study the bandstructure of twisted moir\'e potentials in detail. At sets of commensurate twisting angles, the low energy single-particle spectrum of a twisted moir\'e potential will form into distinct bands and gaps. To a first approximation, energy bands can be qualitatively modelled by harmonic states, localised in different potential minima. The bands are intrinsically linked to the number of distinct minima and size of the moir\'e unit cell, with smaller cells producing larger gaps and vice versa. For shallower potential depths, degeneracies between harmonic states are lifted by virtue of anharmonic confinement and coupling between states. Depending on the exact geometry of potential minima, bands can then be classified by $4$ unique forms of tight-binding models. We find excellent agreement between the continuous spectrum and fitting to our tight-binding models, allowing for accurate tunnelling rates and onsite energies to be extracted. Our results are directly relevant to the bosonic, many-body problem, and thus provide further understanding on the relative stability of quantum phases both in theory and experiments. In particular, the prominence of gaps can be mapped to strongly correlated insulating phases. Furthermore, tunnelling rates of different bands serve as thresholds on temperature in which a phase can be either a normal fluid or superfluid.
\end{abstract}

\maketitle

\begin{figure*}[t!]
	\centering
	\makebox[0pt]{\includegraphics[width=0.99\linewidth]{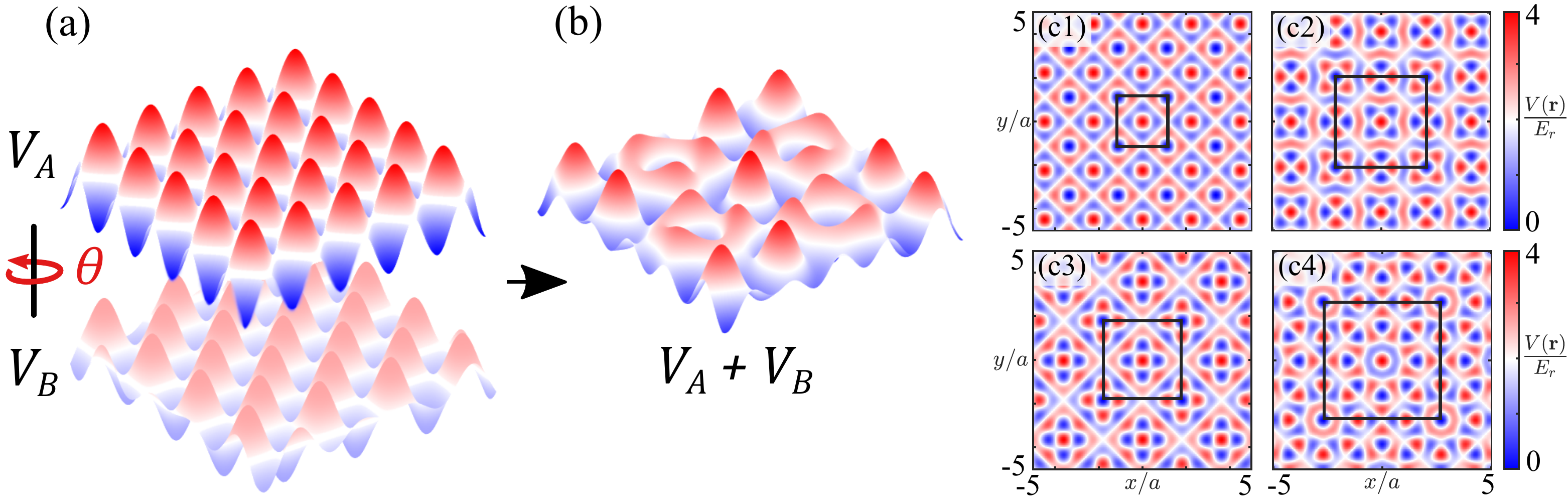}}
	\caption{Illustration of twisted moir\'e potentials. Here, we take (a)~two optical lattices $V_A$ and $V_B$ with a twist angle $\theta$ between them. By superimposing the two lattices, we generate a (b)~twisted moir\'e potential.
	Examples of moir\'e potentials are plotted for commensurate angles (c1)~$\theta_{2,1}\approx36.87^\circ$, (c2)~$\theta_{3,5}\approx28.07^\circ$, (c3)~$\theta_{2,3}\approx22.37^\circ$, and (c4)~$\theta_{2,5}\approx43.60^\circ$, with black squares showing the unit cells.}
	\label{fig_m_potential}
\end{figure*}

\section{Introduction}
In condensed matter physics, twistronics has emerged as an exciting field where novel electronic phenomena are observed~ \cite{Carr2017,Hennighausen_2021}. By stacking layers of twisted, 2D materials, a wealth of exotic features result from the complex interplay of quantum interference, inter-particle interactions, and structural properties, ranging from moir\'e superlattices~\cite{Santos2007,Santos2012,Abbas2020,brzhezinskaya2021engineering}, excitons~\cite{Choi2021,wilson2021excitons,ciarrocchi2022excitonic}, and quantum Hall effects~\cite{Lee2011,Moon2012,tseng2022anomalous}. A key component to the realisation of these properties relates to the underlying, single-particle bandstructure.
For instance, in twisted bilayer graphene, the appearance of flat momentum bands at magic twisting angles significantly enhances quantum correlations~\cite{Morell2010,Bistritzer2011,Zou2018,Tarnopolsky2019,choi2019electronic}, and has led to the appearance of unconventional superconductivity~\cite{cao2018unconventional,Po2018,Gonzalez2019,Yankowitz2019} and highly correlated insulators~\cite{cao2018correlated,Xie2020,saito2020independent}.
In addition, twisted systems allow to interpolate between periodic and quasiperiodic structures by adjusting the twist angle. In some materials, however, inter-layer couplings favour periodic structures, which obscures quasiperiodic effects~\cite{Santos2007,Hass2008,Kim2013}.

Over recent years, ultracold atomic gases have been proposed as highly versatile platforms to quantum simulate a variety of problems~\cite{Lewenstein2007,Bloch2008,bloch2012quantum,gross2017,esslinger2010fermi,tarruell2018}, including non-standard lattice models~\cite{Dutta_2015} and twisted systems~\cite{cirac2019,Salamon2020,meng2023atomic}.
Bilayer models can be emulated, in which different internal atomic states represent the different layers, which are Raman coupled and subjected to optical lattices twisted with respect to each other~\cite{cirac2019}.
Alternatively, a single-layer model can be realised when both twisted optical lattices apply to the same internal state~\cite{johnstone2024weak}.
The latter case also models bilayers in the strong interlayer coupling limit~\cite{meng2023atomic}.
In all cases, the twist angle can be freely tuned in experiments by adjusting the angular alignment of optical lattices.

Generally speaking, arbitrary twist angles result in quasiperiodic, or quasicrystalline, optical potentials. An interesting set of quasiperiodic potentials are those which possess rotational symmetry incompatible with crystalline order, i.e.~superimposed optical lattices that are aligned across forbidden rotational symmetries~\cite{Guidoni1997,lsp2005,jagannathan2013}. In recent years, the exotic properties of these quasiperiodic systems have been extensively studied in both single-particle~\cite{viebahn2019,sbroscia2020,Zhu2024} and many-body~\cite{Johnstone2019,Gautier2021,ciardi2022,Zhu2023,zampronio2024} scenarios. Alternatively, for a countable set  of twist angles, the same devices realise superlattice moir\'e patterns with rich internal structure.
Such potentials
likewise exhibit novel properties that are distinct from ordinary, periodic systems~\cite{Buonsante2005,Jo2012,Chern2014,Thomas2017,johnstone2024weak}.


In both cases, the gapped-band structure plays a central role in determining the physical behaviour of single particles, as well as the many-body properties.
Importantly, bands and subbands can be mapped onto effective tight-binding models, which describe a network of sites associated to localized Wannier states coupled by tunnelling processes.
In the incommensurate case, the strong inhomogeneities of the potential induces a complex set of non-uniform Wannier states, from which such models may be built~\cite{gottlob2023}.
In the commensurate case, one can take advantage of periodicity.
For simple optical potentials such as a square lattice, each site is associated to a potential minimum, and the structure of the lattice simply reproduces the Bravais lattice of elementary cells.
In contrast, for moir\'e potentials, each elementary cell has several potential minima that are not equivalent, and effective tight-binding models need to be constructed with care.

An approach to constructing effective tight-binding models for optical moir\'e optical potentials has been outlined in a previous work~\cite{johnstone2024weak}.
The primary objective of the present paper is to discuss this construction in more detail and extract effective tight-binding parameters by fitting to the general, continuous-space model.
We find that the presence of several sets of potential minima in each unit cell induces the splitting of energy bands into subbands. Each band is associated to a different, effective tight-binding model, some of which possess a structure that differs from the initial Bravais lattice.
Moreover, effective tunnelling rates are associated to the width of narrow subbands, which significantly lowers the energy scale where long-range quantum coherence is to be expected.
%
%

The layout of our results is as follows. First, we introduce twisted optical moir\'e potentials in Sec.~\ref{section_model}, alongside a discussion of the spectral properties as obtained from the exact continuous-space model in Sec.~\ref{section_spect}. Comparison to a harmonic approximation of the moir\'e potential in the vicinity of the various minima allows us to classify the main bands. We then discuss in Sec.~\ref{section_tbm} the construction of effective tight-binding models in each band, taking advantage of energy-scale separation between intra-cell and inter-cell couplings. The behaviour of the effective tight-binding parameters obtained from fitting to exact continuous-space calculations is discussed alongside the validity of the tight-binding models in Sec.~\ref{section_ft}. We finally discuss extension to other moir\'e potentials in Sec.~\ref{section_general}, before ending with our conclusions in Sec.~\ref{section_cnc}.

\section{Twisted Moir\'e Potentials} \label{section_model}
We consider single-particles trapped in a twisted optical lattice $V(\mathbf{r})$, with Hamiltonian
\begin{equation} \label{eq_sph}
\begin{aligned}
\hat{H} = -\dfrac{\hbar^2 \mathbf{\nabla}^2}{2M} + V(\mathbf{r}),
\end{aligned}
\end{equation}
where $M$ is the particle mass and $\mathbf{r}=(x,\, y)$ is the position. The twisted optical lattice is defined by
\begin{equation} \label{eq_moireL}
V(\mathbf{r}) = V \left[ v(R^{+} \mathbf{r}) + v(R^{-} \mathbf{r}) \right],
\end{equation}
where
\begin{equation}
\begin{aligned}
v(\mathbf{r}) = \cos^2 (\pi x/a) + \cos^2 (\pi y/a),
\end{aligned}
\end{equation}
$V$ is the potential depth, $R^{\pm}$ is the rotation matrix with angle $\pm \theta/2$, $a=\lambda/2$ is the lattice constant, and $\lambda$ is the optical wavelength. All energies will be expressed in terms of the recoil energy $\Er=\pi^2 \hbar^2/2Ma^2$. A visualisation of the potential is given in Figs.~\ref{fig_m_potential}(a)-(b).
For a specific set of twist angles $\theta$, known as commensurate angles, it is possible to form periodic, moir\'e patterns in real space.
Here we only outline their properties, with further details and proofs presented in Appendix~\ref{app_geo}.
The commensurate, or moir\'e angles, can be directly written as
\begin{equation} \label{eq_c_angle}
\theta_{m,n} = \cos^{-1}\Big(\frac{2mn}{m^2+n^2}\Big),
\end{equation}
where $m$ and $n$ are coprime integers~\cite{cirac2019}.
Since the underlying structure is periodic and shows four-fold rotation symmetry, the system possess a well-defined, square unit cell with dimensionless size
\begin{equation}	\label{eq_l_length}
\ell_{m,n}/a = \begin{cases}
\sqrt{(m^2 + n^2)/2},& \textrm{if } m+n\, \textrm{even},\\
\sqrt{m^2 + n^2},& \textrm{if } m+n \, \textrm{odd}.
\end{cases}
\end{equation}
This set of moir\'e angles can be deduced by considering the intersection of two lattice vectors tilted by $\theta_{m,n}$, where $m$ and $n$ define a point of intersection at $\mathbf{r}/a=(m,n)$. To determine the true period, it is important to note that geometrically equivalent moir\'e angles can exist for different $m$ and $n$, hence the distinction between $m+n$ being even or odd, see details and proof in Appendix~\ref{app_geo}.
In Figs.~\ref{fig_m_potential}(c1)-(c4), some examples of the moir\'e potentials are plotted, alongside their corresponding moir\'e unit cells. For larger integers $m$ and $n$, the underlying moir\'e cell is enlarged, with richer inhomogeneous structure. As will be shown later in our discussions, a very important quantity that characterises the bandstructure is the total number of distinct, local minima/maxima in the potential, which is given by
\begin{equation}	\label{eq_m}
\mathcal{M}_{m,n} = \begin{cases}
(\ell^2_{m,n}/a^2 -1)/4,& \textrm{if } m+n\, \textrm{even},\\
(\ell^2_{m,n}/a^2 -1)/4 + 1,& \textrm{if } m+n \, \textrm{odd}.
\end{cases}
\end{equation}
Note, for this equation, $\mathcal{M}_{m,n}$ is always an integer, with a formal proof provided in Appendix~\ref{app_geo}.

\section{Spectral Properties} \label{section_spect}

\subsection{Dispersion Relations}\label{sec:Dispersion}
For a periodic potential such as $V(\mathbf{r})$, the Hamiltonian $\hat{H}$ my be solved using a standard Bloch transformation~\cite{ashcroft1976}.
We write the
wavefunction as $\psi (\mathbf{r})=e^{i\mathbf{k} \cdot \mathbf{r}} u(\mathbf{r})$,
where $u(\mathbf{r})$ is a Bloch function, which is periodic with period equal to the moir\'e unit length $\ell_{m,n}$,
and $\mathbf{k}=(k_x, k_y)$ is the quasi-momentum, which can be restricted to the 1st Brillouin zone, $k_{x,y} \ell_{m,n} \in [-\pi \dots \pi]$.
Inserting the Bloch transformed wavefunction into the Schr\"odinger equation $\hat{H}\psi (\mathbf{r}) = E\psi (\mathbf{r})$,
we obtain the reduced equation
\begin{equation} \label{eq_sph_k}
\begin{aligned}
\varepsilon(\mathbf{k}) u(\mathbf{r}) = \left[ \dfrac{\hbar^2}{2M}\left( \mathbf{k}^2 - 2 i \mathbf{k} \cdot \nabla - \nabla^2 \right) + V(\mathbf{r})\right]u(\mathbf{r}),
\end{aligned}
\end{equation}
where $\varepsilon(\mathbf{k})$ spans the set of energies $E$ that fulfill the periodic boundary conditions for each quasi-momentum $\mathbf{k}$.
We then solve for the Bloch functions $u(\mathbf{r})$ and eigenenergies/dispersion relations $\varepsilon(\mathbf{k})$ via exact diagonalisation, using discretisation with a grid spacing of at least $h/a = 0.05$. Since $u(\mathbf{r})$ is periodic in the moir\'e unit cell, we can therefore diagonalise in an $\ell_{m,n} \times \ell_{m,n}$ box with periodic boundary conditions without loss of generality.

\begin{figure}[t!]
	\centering
	\makebox[0pt]{\includegraphics[width=1.0\linewidth]{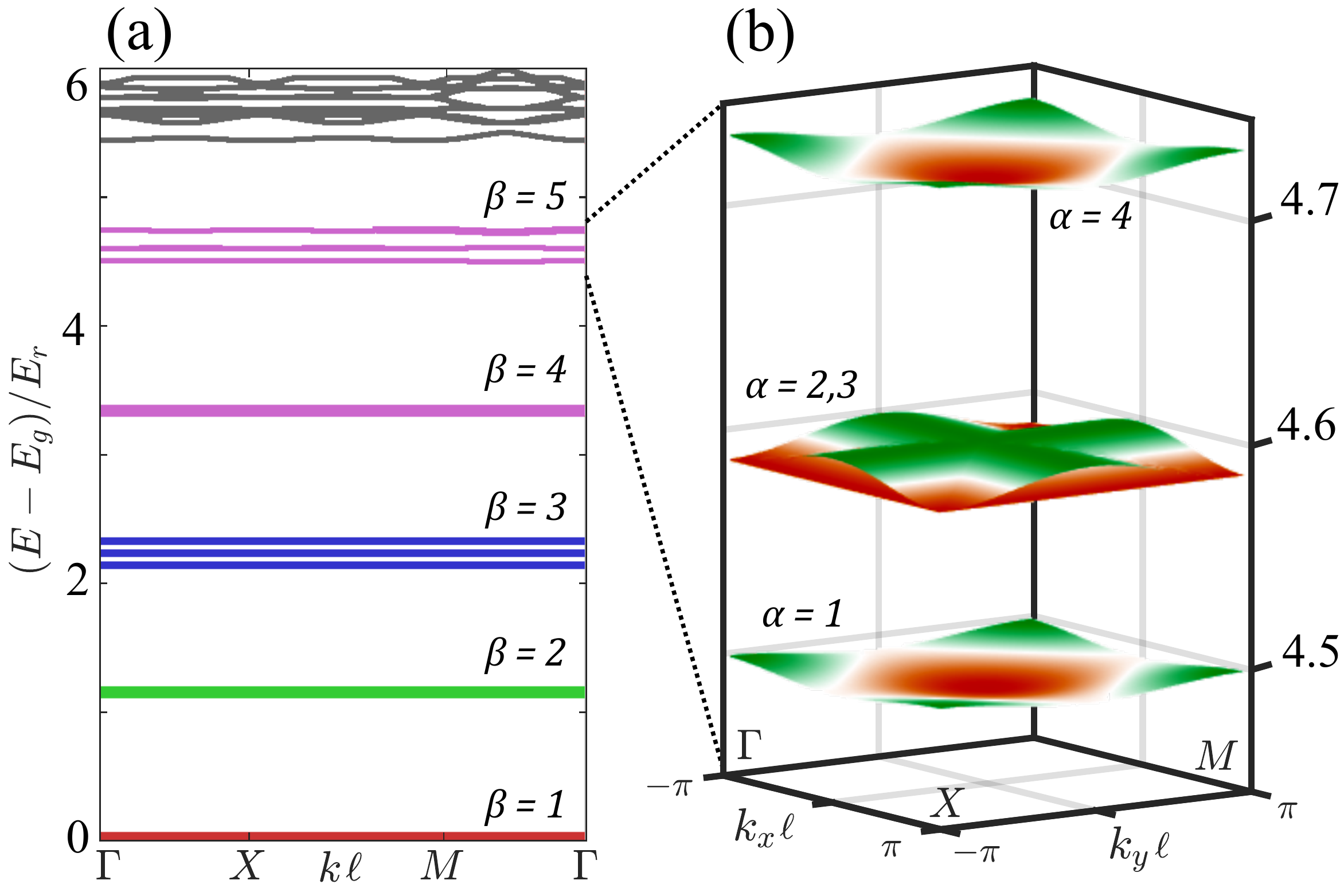}}
	\caption{Exact energy spectrum  $(E-E_g)/\Er$, with $E_g$ the single-particle ground state energy, obtained from the continuous model for amplitude $V=6\Er$ and twist angle $\theta_{3,5}\approx28.07^\circ$. Coloured lines denote different bands of states with index $\beta$, with colours representing distinct minima over which a state is distributed, see~Fig.~\ref{fig_harmonicC_min}(a). In~(a), we plot the dispersion relations across high-symmetry points of the first Brillouin zone, with a zoom in~(b) given to the subbands with index $\alpha$ in band $\beta=5$. For $E-E_g \gtrsim 5\Er$, well-separated bands can no longer be identified (grey lines).}
	\label{fig_disp}
\end{figure}

In Fig.~\ref{fig_disp}, we plot an example of the dispersion relations $\varepsilon(\mathbf{k})$ hence obtained in the 1st Brillouin zone for the moir\'e potential with twist angle $\theta_{3,5}\approx28.07^\circ$ and amplitude $V=6\Er$.
As expected for a sufficiently deep potential, we observe the formation of wide spectral gaps, separating almost flat dispersive bands, see Fig.~\ref{fig_disp}(a).
The different bands are labelled with an index $\beta$, which ranges from $1$ to $5$ for the considered energy range.
Each band (except $\beta=1$) splits into a set of narrower subbands (with index $\alpha$). A magnification of band $\beta=5$ plotted in Fig.~\ref{fig_disp}(b),
shows that each subband displays a cosine-like dispersion, which is to be expected for a deep optical lattice.
Note that the subbands $\alpha=2$ and $\alpha=3$ are quasi-degenerate, each one corresponding to one branch of the observed cross structure.
Qualitatively similar properties are observed for the other bands and at other moir\'e angles.

The main bands ($\beta$) are explained by a simple harmonic approximation around the local minima of the moir\'e potential as discussed in Sec.~\ref{section_harmonic}.
In contrast, the subbands ($\alpha$) and their dispersion relations are explained by tight-binding models, the structure of which depends on the band, as discussed in Sec.~\ref{section_tbm}.

\begin{figure}[t!]
	\centering
	\makebox[0pt]{\includegraphics[width=0.99\linewidth]{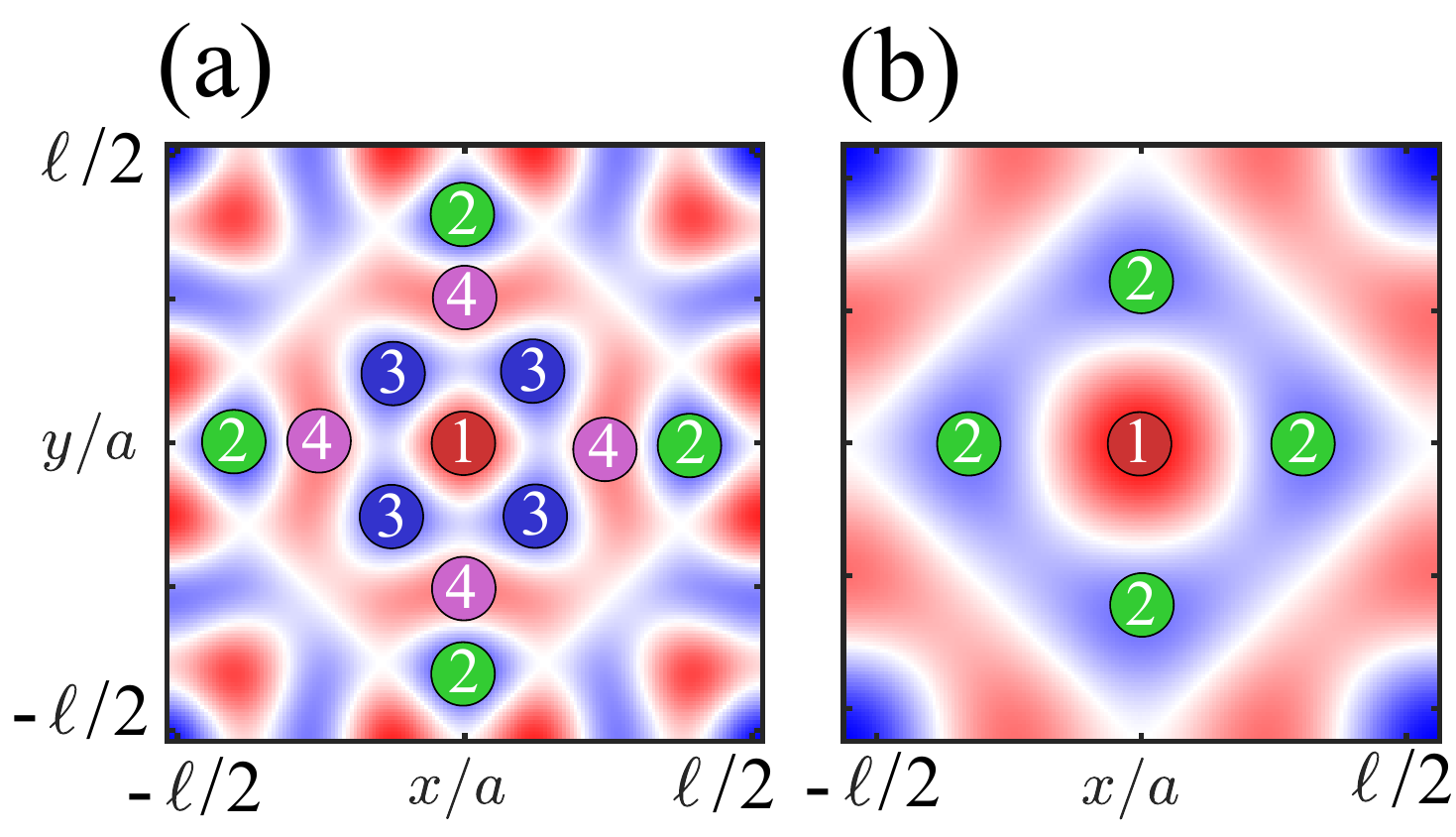}}
	\caption{Zoom of two twisted moir\'e unit cells for angles (a)~$\theta_{3,5}\approx28.07^\circ$, where $\mathcal{M}_{3,5}=4$,
	and (b)~$\theta_{2,1}\approx36.87^\circ$, where $\mathcal{M}_{2,1}=2$, using the same colourscale as Fig.~\ref{fig_m_potential}. Coloured points denote distinct sets of potential minima with index $u$ (white numbers), ordered from lowest to highest energy. Note, minima located on the unit cell edges/corners are marked by their geometrically identical local maxima within the unit cell (sets 1 and 4), for clarity (see text).
	}
	\label{fig_harmonicC_min}
\end{figure}

\begin{figure*}[t!]
	\centering
	\makebox[0pt]{\includegraphics[width=0.99\linewidth]{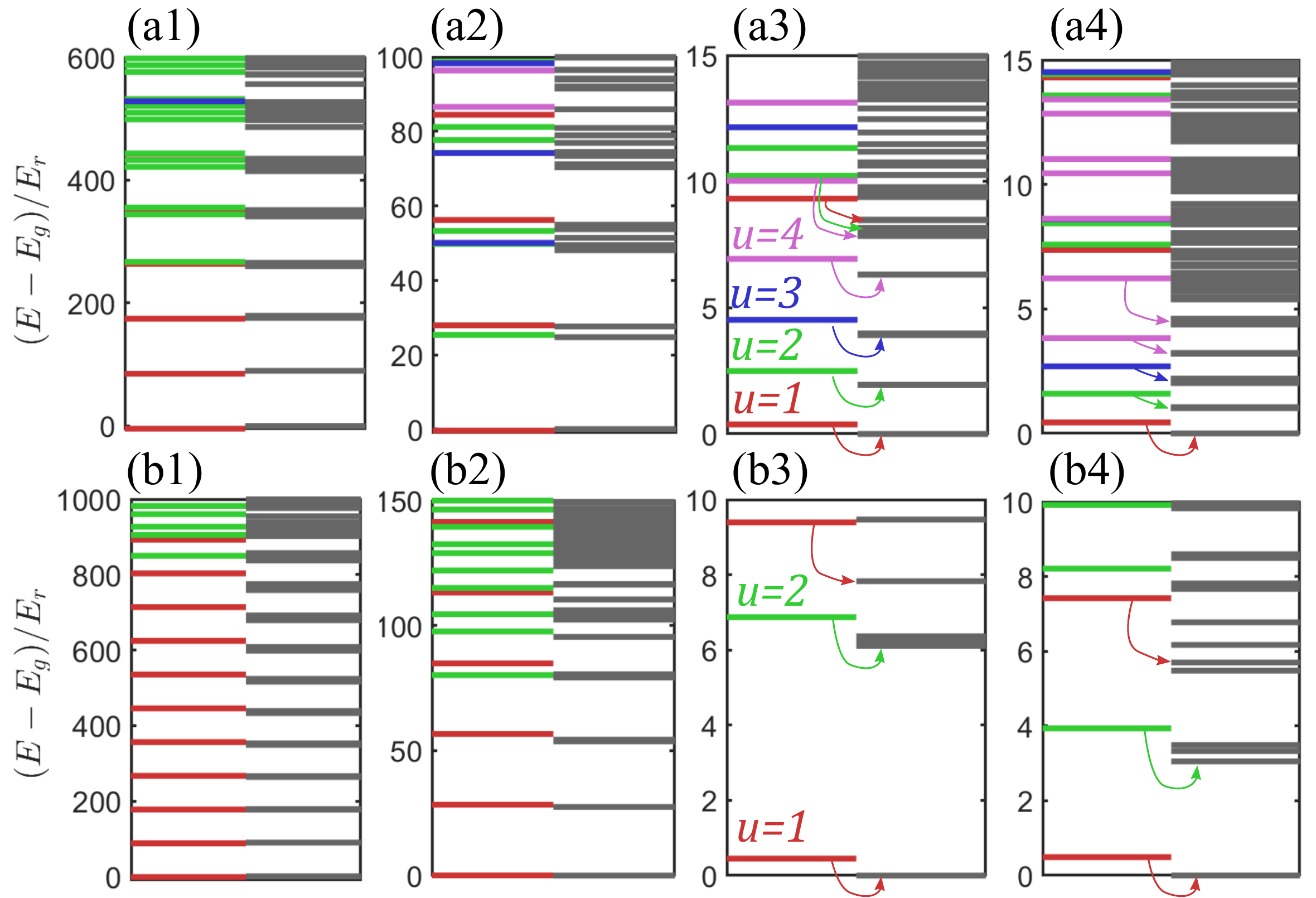}}
	\caption{Comparison of continuous (grey lines) and harmonic (coloured lines) energy spectra $(E-E_g)/\Er$, where $E_g$ is the continuous ground state energy. The lowest grey lines thus correspond to $E_g=0$.
	The upper line, (a1)-(a4), corresponds to the twist angle $\theta_{3,5}\approx28.07^\circ$, the second line, (b1)-(b4), to $\theta_{2,1}\approx36.87^\circ$.
	From the left to right columns, we consider decreasing potential depths of (a1),(b1)~$V=1000\Er$, (a2),(b2)~$V=100\Er$, (a3),(b3)~$V=10\Er$, and (a4),(b4)~$V=6\Er$.
	The colours of harmonic energies denote the $u$-th distinct set of potential minima, as per Fig.~\ref{fig_harmonicC_min}, with some arrows at lower $V/E_r$ indicating bands in which continuous-space eigenstates match/look similar to a harmonic eigenstate.}
	\label{fig_harmonicC}
\end{figure*}


\subsection{Harmonic Approximation} \label{section_harmonic}

If $V/\Er$ is sufficiently large, the eigenstates are distributed around the local minima of the potential. Neglecting tunnelling between degenerate local minima, which are sufficiently far apart, each local minimum may accommodate a set of 
well-localised Wannier functions, corresponding to the ground and excited states in a given well.
Figures~\ref{fig_harmonicC_min}(a) and (b) show such local minima in a unit cell for the moir\'e potentials with twist angles $\theta_{3,5}\approx28.07^\circ$ and $\theta_{2,1}\approx36.87^\circ$, respectively. The lowest minimum (red, $u$=1) is unique, while the higher ones (green, $u=2$; blue, $u=3$; purple, $u=4$) are four-fold degenerate. Note that, according to Eqs.~(\ref{eq_l_length}) and (\ref{eq_m}), there are, respectively, $\mathcal{M}_{3,5}=4$ and $\mathcal{M}_{2,1}=2$ distinct local minima in each unit cell. By shifting V($\mathbf{r}$) such that the global minimum is enclosed in the unit cell centre, e.g. a positional shift by $(+\ell/2, \, +\ell/2)$ and setting $+V$ to $-V$, we have the same potentials as depicted in Fig.~\ref{fig_harmonicC_min}. In other words, twisted square moir\'e potentials will exhibit the same physics for both blue- or red-detuned systems (i.e. localisation of atoms to minima or maxima), hence we write the minima with $u=1$ (red) and $u=4$ (purple) at the equivalent sets of local maxima in Fig.~\ref{fig_harmonicC_min} for compactness.
Expanding the potential around a local minimum centred at $\mathbf{R}$, we write
\begin{eqnarray} \label{eq_Vt}
V(\mathbf{R}+\mathbf{r}) & = &V(\mathbf{R}) + \frac{1}{2} \big(\Omega_{xx} x^2 + 2\Omega_{xy} x y +\Omega_{yy} y^2 \big)
\nonumber \\ 
& & + \mathcal{O}(\dots),
\end{eqnarray}
where $\Omega_{uv} = \partial^2 V/\partial u \partial v$
and $\mathcal{O}(\dots)$ accounts for anharmonic corrections.
Diagonalizing the Hessian matrix $\Omega$, we then find
\begin{equation} \label{eq_VtF}
\begin{aligned}
V(\mathbf{R}+\mathbf{r}) \approx V(\mathbf{R}) + \frac{M}{2} \left(\omega^2_{+} x'^2 + \omega^2_{-} y'^2 \right),
\end{aligned}
\end{equation}
with
\begin{equation}
\begin{aligned}
\omega_\pm^2 = \frac{\Omega_{xx} + \Omega_{yy} \pm \sqrt{(\Omega_{xx} - \Omega_{yy})^2 + 4\Omega_{xy}^2}}{2M},
\end{aligned}
\end{equation}
and $x'$-$y'$ coordinates in a rotated orthogonal frame.
The eigenvalues of the 2D harmonic oscillator in Eq.~\eqref{eq_VtF} are given by
\begin{equation} \label{eq_harmonic_E}
\begin{aligned}
E_{n_+,n_-} = V(\mathbf{R}) + \hbar \omega_+ \left( n_+ + 1/2\right) + \hbar\omega_- \left(n_- + 1/2\right),
\end{aligned}
\end{equation}
where $n_\pm \in \mathbb{N}$.
If $\omega_{+}=\omega_{-}$, the $n$-th excited state of the spectrum will be $(n+1)$-fold degenerate.
However, in general we have $\omega_{+} \neq \omega_{-}$, i.e. all degeneracies are lifted, except for some accidental matchings.
The final spectrum is then the combination of all energies from Eq.~\eqref{eq_harmonic_E} for each potential minimum labelled by its position $\mathbf{R}$.
Each minimum forms a ladder set, which are degenerate within a given family of minima $u$.
Depending on the relative strength of $V/\Er$, different sets for different minima may be located between one another, or potentially overlap, leading to an intricate spectrum.

In Figs.~\ref{fig_harmonicC}(a1)-(a4), we compare the exact spectra of Eq.~(\ref{eq_sph_k}) (grey lines) at mid band $\mathbf{k}\ell_{m,n}=(\pi/2,\pi/2)$ to that of Eq.~\eqref{eq_harmonic_E}, using the moir\'e angle $\theta_{3,5}$ and for decreasing values of $V/\Er$. Coloured lines show the harmonic energies of Eq.~\eqref{eq_harmonic_E}, with different colours, corresponding to each unique type of potential minimum shown in Fig.~\ref{fig_harmonicC_min} and labelled by $u=1$, $2$, $3$, and $4$. Starting with the larger values of $V/\Er=1000$ and $V/\Er=100$ in Figs.~\ref{fig_harmonicC}(a1)-(a2), we find good agreement between the exact and harmonic spectra. Each minimum forms a series of distinct bands, with the total number of states/degeneracies in each harmonic band matching the number of states in continuous-space bands. Prominent gaps can also form, depending on the relative separation between eigenenergies of the local minima. Discrepancies and degeneracy lifting between exact and harmonic eigenenergies becomes more apparent at higher energy. This is to be expected, since the higher energy eigenstates contain larger components further away from a local minima, and anharmonic corrections
become more important.
For smaller lattice amplitudes, for instance $V/\Er=10$ and $V/\Er=6$ in Figs.~\ref{fig_harmonicC}(a3)-(a4), we still find some qualitative agreement between the exact and harmonic results for some of the lowest energy states. Some coloured arrows for the lowest energy states are also plotted, indicating which continuous-space bands match with the harmonic bands in terms of similar looking eigenstates. Discrepancies between energy levels are again seen, due to the fact that for smaller $V/\Er$, low energy eigenstates also have a larger extent, i.e. anharmonic terms are non-negligible.
Overall, we find that the low energy bands and gap structure of the twisted moir\'e potential remains reminiscent to that of the harmonic spectrum, with quantitative accuracy for larger potential depths.


Equivalent properties are also observed at different moir\'e angles, e.g.~$\theta_{2,1}$ as shown in Figs.~\ref{fig_harmonicC}(b1)-(b4). For this angle, we generally find larger gaps compared to $\theta_{3,5}$, which can be explained as follows. Each $\theta_{m,n}$ corresponds to $\mathcal{M}_{m,n}$ distinct local minima. If $\mathcal{M}_{m,n}$ is small (i.e.~the moir\'e cell is small), the relative energy between different minima will be large, therefore producing wide spectral gaps. However, if $\mathcal{M}_{m,n}$ is large (i.e.~the moir\'e cell is large), the underlying gaps become less significant due to the smaller relative energy difference between minima. 


\section{Tight-Binding Models} \label{section_tbm}
As discussed in the previous sections, the single-particle spectrum of twisted optical potentials separates into distinct energy bands with prominent gaps between them, see Fig.~\ref{fig_disp}.
The central energy of the latter can be approximately understood via harmonic bands associated to different sets of local minima.
In order to describe the dispersion relation of the various bands, we now need to introduce finite tunnel couplings between these minima.
Tunnel processes are dominated by the resonant ones, i.e~those that couple minima with equal energies.
This leads to tight-binding models with different structures in different bands, as we discuss now.
%
Throughout this section, we focus on the case of amplitude $V=6\Er$ and twist angle $\theta=\theta_{3,5}$ from Fig.~\ref{fig_disp}.
However, as we will show, the approaches and classifications here can be applied to any commensurate angle and sufficiently large potential amplitude.
To ease on notation, we omit the commensurate twist angle indices and write the moir\'e unit length $\ell=\ell_{m,n}$.

\subsection{Band $\beta=1$} \label{sec_tbm_b1}
\begin{figure}[t!]
	\centering
	\makebox[0pt]{\includegraphics[width=0.99\linewidth]{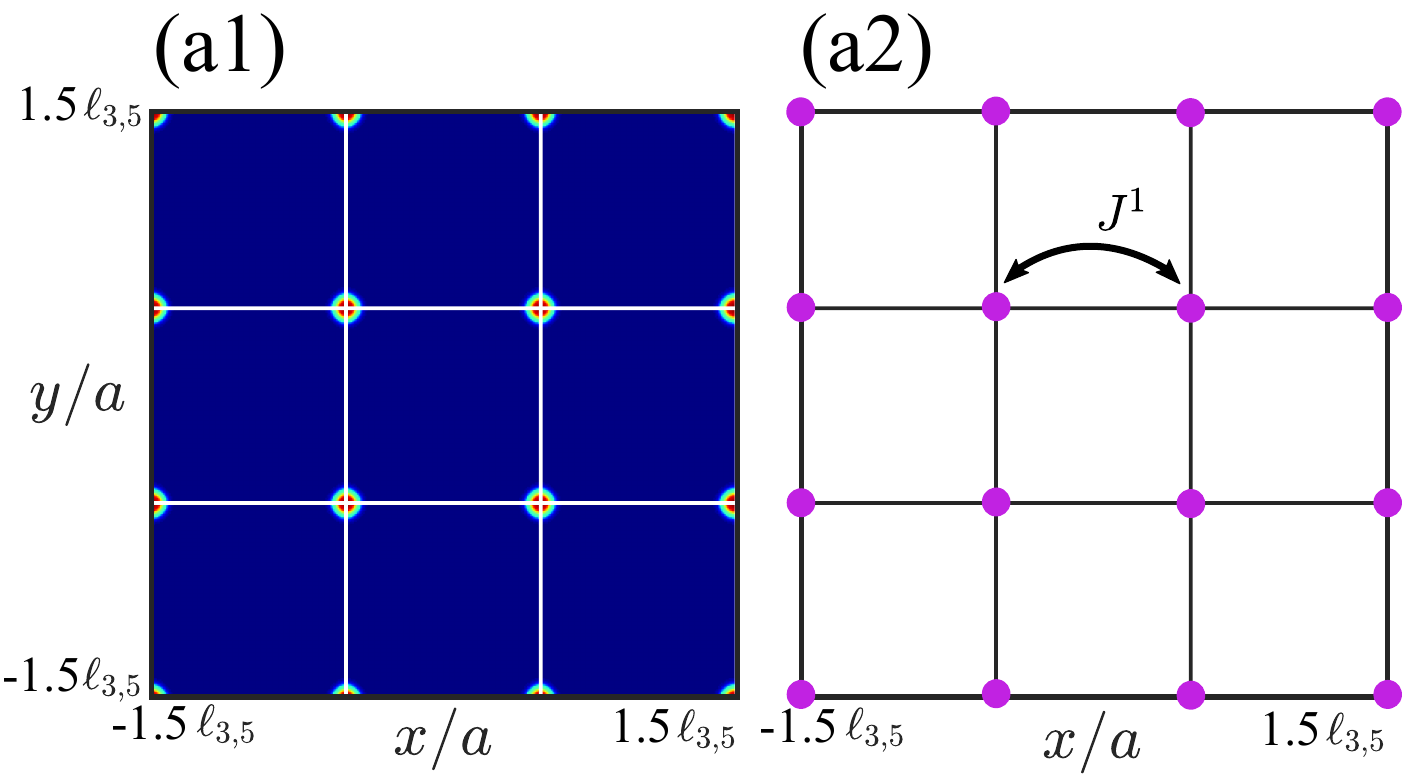}}
	\caption{Plot of the (a1)~ground state density profile for band~$1$ and (a2)~the corresponding tight-binding model, taking density spots as lattice sites (purple circles) with nearest-neighbour coupling $J^1$. This produces a square lattice with period $\ell$, where the white/black squares in (a1)/(a2) denote the moir\'e unit cell.}
	\label{fig_band1}
\end{figure}

We start with simplest case of band $1$, which describes the ground state, the density distribution of which is shown in Fig.~\ref{fig_band1}(a1).
The approach we use is standard and we briefly outline it for reference.
Since there is a single minimum in each unit cell, we can model the band with a simple square lattice, with sites located at those minima, as per Fig.~\ref{fig_band1}(a2). The Hamiltonian is
\begin{equation} \label{eq_h_b1}
\hat{H} = \epsilon^\beta\sum_{i} \hat{a}^\dagger_i \hat{a}_i -J^\beta\sum_{\langle i,j \rangle} \hat{a}^\dagger_i \hat{a}_j,
\end{equation}
where $i,j$ are site indices, $\epsilon^\beta$ is the onsite energy in band $\beta=1$, $J^\beta$ is the corresponding tunnelling between sites, and $\langle i,j \rangle$ denotes summation over nearest-neighbouring sites.
The dispersion relation is readily found by introducing the Fourier transformed operators
\begin{equation} \label{eq_ft_a}
\hat{a}_i=\frac{1}{\sqrt{N}}\sum_\mathbf{k} \hat{a}_\mathbf{k} e^{i \mathbf{k} \cdot \mathbf{r}_i},
\end{equation}
where $N$ is the total number of sites and $\mathbf{k}$ spans the first Brillouin zone, i.e. $k_{x,y} \ell\in [-\pi \dots \pi]$.
Using
\begin{equation}
\sum_{\mathbf{k},\mathbf{k}',i} e^{i (\mathbf{k}-\mathbf{k}') \cdot \mathbf{r}_i} = N \delta_{\mathbf{k},\mathbf{k}'}
\end{equation}
and $\mathbf{r}_i$
spanning the lattice sites, Hamiltonian~\eqref{eq_h_b1} may be written as
\begin{equation}
\hat{H} = \sum_\mathbf{k} \varepsilon(\mathbf{k}) \hat{a}^\dagger_\mathbf{k} \hat{a}_\mathbf{k},
\end{equation}
where the dispersion relation $\varepsilon(\mathbf{k})$ is given by
\begin{equation} \label{eq_k_sq}
\varepsilon(\mathbf{k})= \epsilon^\beta - 2J^\beta(\cos k_x \ell + \cos k_y \ell ).
\end{equation}
This generates a standard band dispersion relation with cosine dependence in both $x$ and $y$ directions.
The unknown quantities $\epsilon^\beta$ and $J^\beta$ can then be readily extracted by fitting Eq.~(\ref{eq_k_sq}) to the exact dispersion relation found from continuous space calculations as done in Sec.~\ref{sec:Dispersion}.
The result is discussed in Sec.~\ref{section_ft}.

\begin{figure}[t!]
	\centering
	\makebox[0pt]{\includegraphics[width=0.99\linewidth]{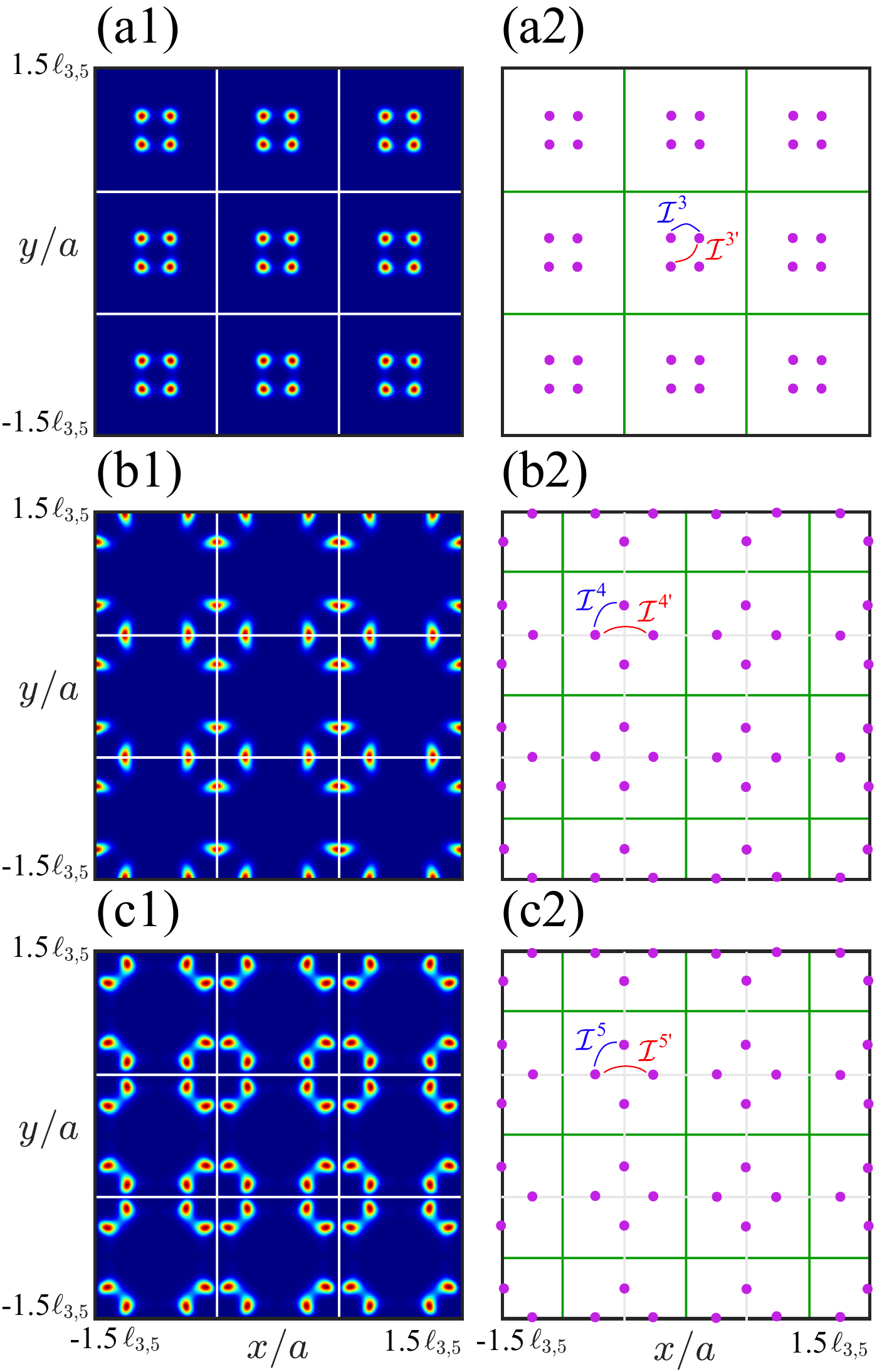}}
	\caption{Plots of the (a1),(b1),(c1)~density profiles for a state in bands $3$, $4$, and $5$, respectively, and (a2),(b2),(c2)~the corresponding tight-binding models, taking density spots (maxima) as lattice sites (purple circles). The green squares in (a2),(b2),(c2) are unit cells of length $\ell$, which contain $4$ sites with nearest-neighbour coupling $\mathcal{I}^\beta$ (blue) and next-nearest-neighbour coupling $\mathcal{I}^{\beta'}$ (red).}
	\label{fig_band345}
\end{figure}

\subsection{Bands $\beta=3,4,5$} \label{sec_tbm_b345}
For all other bands, we have a different situation, where there are now $4$ potential minima within each moir\'e cell, with different geometries in different bands, see~Fig.~\ref{fig_band345} for bands $\beta=3$, $4$, and $5$.
For now, we forgo the discussion of band $2$, which is more complicated, and focus on bands $3$-$5$.
In order to model one of these bands, we may separate the tunnelling rates into two distinct energy scales:
intra-cell tunnellings ($\mathcal{I}^\beta$ and $\mathcal{I}^{\beta'}$ for band $\beta$) and 
inter-cell tunnelling ($J^\beta_\alpha$ for subband $\alpha$ in band $\beta$).
Owing to exponential decay of tunnelling rates with site separation, the intra-cell couplings generally exceed the inter-cell couplings, i.e. $\mathcal{I}^\beta, \mathcal{I}^{\beta'} \gg J^\beta_\alpha$.
We may then treat the inter-cell couplings in perturbation of the intra-cell ones.

We begin by writing the Hamiltonian of an isolated moir\'e cell, i.e.~in one of the green squares in Figs.~\ref{fig_band345}(a2),(b2),(c2). 
Note, in Figs.~\ref{fig_band345}(b2) and (c2), we have shifted the green unit cell across the diagonal in order to enclose $4$ sites in the vicinity of the cell center and have a similar description to band $3$ in Fig.~\ref{fig_band345}(a2). 
Each unit cell contains $4$ sites located at the spots of the considered band. In matrix form, the Hamiltonian can be written as
\begin{equation} \label{eq_h_lt}
\hat{H}_{cell} = 
\begin{pmatrix}
\epsilon^\beta & -\mathcal{I}^\beta & -\mathcal{I}^{\beta'} & -\mathcal{I}^\beta \\
-\mathcal{I}^\beta & \epsilon^\beta & -\mathcal{I}^\beta & -\mathcal{I}^{\beta'} \\
-\mathcal{I}^{\beta'}  & -\mathcal{I}^\beta  & \epsilon^\beta & -\mathcal{I}^\beta  \\
-\mathcal{I}^\beta & -\mathcal{I}^{\beta'} & -\mathcal{I}^\beta & \epsilon^\beta 
\end{pmatrix},
\end{equation}
where $\mathcal{I}^\beta$ denotes nearest-neighbour tunnelling, $\mathcal{I}^{\beta'}$ next-nearest-neighbour tunnelling, and $\epsilon^\beta$ is the onsite energy of the band. Since all sites lie in equivalent potential minima and the system has $4$-fold rotational symmetry, $\epsilon^\beta$ is the same for each site. Moreover, each set of intra-cell tunnellings are equal. The matrix in Eq.~\eqref{eq_h_lt} can then be diagonalised, with normalised eigenstates given by
\begin{equation} \label{eq_eigVec}
\begin{aligned}
\ket{a} = \frac{1}{2}
\begin{pmatrix}
1 \\ 1 \\ 1 \\ 1 
\end{pmatrix}, \, \,
\ket{B} = \frac{1}{\sqrt{2}}
\begin{pmatrix}
1 \\ 0 \\ -1 \\ 0 
\end{pmatrix}, \, \\
\ket{C} = \frac{1}{\sqrt{2}}
\begin{pmatrix}
0 \\ 1 \\ 0 \\ -1 
\end{pmatrix}, \, \,
\ket{d} = \frac{1}{2}
\begin{pmatrix}
1 \\ -1 \\ 1 \\ -1 
\end{pmatrix},
\end{aligned}
\end{equation}
and eigenvalues
\begin{equation}\label{eq:EigenVal}
\begin{aligned}
E^\beta_a & = \epsilon^\beta + 2\mathcal{I}^\beta -\mathcal{I}^{\beta'}, \\ E^\beta_{B,C} & = \epsilon^\beta + \mathcal{I}^{\beta'}, \\ E^\beta_d & = \epsilon^\beta -2\mathcal{I}^\beta -\mathcal{I}^{\beta'},
\end{aligned}
\end{equation}
where the degeneracy $E^\beta_B=E^\beta_C$ also arises due to the $4$-fold rotational symmetry.
Each isolated moir\'e cell of Fig.~\ref{fig_band345} can thus be described as a $4$-level system, with each level corresponding to each eigenstate of Hamiltonian~\eqref{eq_h_lt}.
Two of them ($a$ and $d$) are isolated in energy while two other ones ($B$ and $C$) are exactly degenerate due to a fundamental symmetry of the system.
We adopt a convention where capital letters denote degenerate states and lower-case non-degenerate states.

\begin{figure}[t!]
	\centering
	\makebox[0pt]{\includegraphics[width=0.99\linewidth]{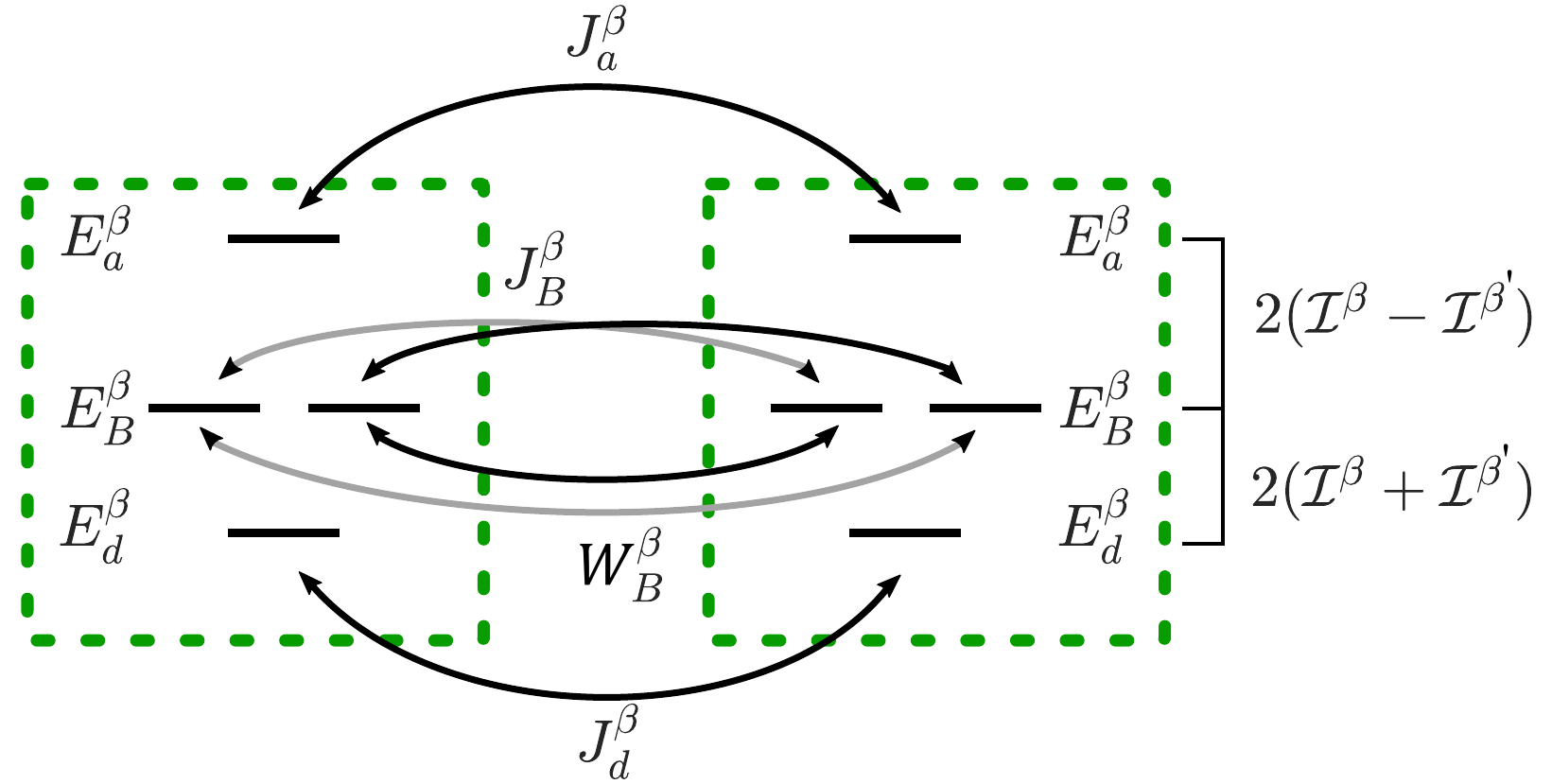}}
	\caption{Inter-cell couplings between the intra-cell eigenstates from Eq.~\eqref{eq_eigVec} from nearest-neighbour green unit cells in Fig.~\ref{fig_band345}(a2),(b2),(c2). Black lines denote the states of each subband $\alpha$, with corresponding energy $E_\alpha^\beta$, where $E_B^\beta$=$E_C^\beta$. The energy spacing between states is shown outside the righthand green cell.}
	\label{fig_band3_coupling}
\end{figure}

To capture tunnelling between nearest-neighbour moir\'e cells, we then couple the $4$-level systems by different inter-cell tunnellings, as depicted in Fig.~\ref{fig_band3_coupling}.
This forms an effective square lattice with period $\ell$, where each site has $4$ internal states.
Restricting ourselves to the dominant (resonant) inter-cell tunnelings between equal energy eigenstates, we find the effective tight-binding Hamiltonian
\begin{equation} \label{eq_h_lt_full}
\begin{aligned}
\hat{H} = & E^\beta_a\sum_i \hat{a}_i^\dagger \hat{a}_i - J^{\beta}_a \sum_{\langle i,j \rangle} \hat{a}_i^\dagger \hat{a}_j + E^\beta_B\sum_i (\hat{B}_i^\dagger \hat{B}_i + \hat{C}_i^\dagger \hat{C}_i)   \\
&-J^{\beta}_B \sum_{\langle i,j \rangle} ( \hat{B}_i^\dagger \hat{B}_j + \hat{C}_i^\dagger \hat{C}_j )  - \sum_{\langle i,j \rangle} W^{\beta}_{B;i,j} ( \hat{B}_i^\dagger \hat{C}_j + \hat{C}_i^\dagger \hat{B}_j )  \\
& + E^\beta_d\sum_i \hat{d}_i^\dagger \hat{d}_i -J^{\beta}_d \sum_{\langle i,j \rangle} \hat{d}_i^\dagger \hat{d}_j,
\end{aligned}
\end{equation}
where $\hat{a}_i$, $\hat{B}_i$, $\hat{C}_i$, and $\hat{d}_i$ are the annihilation operators of a particle in site $i$ for the corresponding eigenstate in Eq.~\eqref{eq_eigVec}.
Note, a directional dependence in the coupling $W^{\beta}_{B;i,j}$ between different degenerate states from adjacent sites is necessary to account for the crossed features of quasi-degenerate subbands as in the centre of Fig.~\ref{fig_disp}(b).
However, overall 4-fold symmetry of the system implies that the values of the coefficients $W^{\beta}_{B;i,j}$ are opposite in orthogonal directions,
i.e.~$W^{\beta'}_{B;x}=-W^{\beta}_{B;y}$, see details in Appendix~\ref{app_latticeIII}.
To determine the dispersion relations, we again introduce Fourier transform operators as in Eq.~\eqref{eq_ft_a} for each state $\ket{a}$, $\ket{B}$, $\ket{C}$, and $\ket{d}$. The momentum space Hamiltonian is then
\begin{equation} \label{eq_h_ft}
\begin{aligned}
\hat{H} = & \sum_\mathbf{k} \varepsilon^\beta_a(\mathbf{k}) \hat{a}_\mathbf{k}^\dagger \hat{a}_\mathbf{k} + \sum_\mathbf{k} \varepsilon^\beta_d(\mathbf{k}) \hat{d}_\mathbf{k}^\dagger \hat{d}_\mathbf{k} \\
& + \sum_\mathbf{k} \varepsilon^\beta_B(\mathbf{k}) ( \hat{B}_\mathbf{k}^\dagger \hat{B}_\mathbf{k} + \hat{C}_\mathbf{k}^\dagger \hat{C}_\mathbf{k} ) \\
& + \sum_\mathbf{k} \varepsilon^{\beta'}_B(\mathbf{k}) (\hat{C}_\mathbf{k}^\dagger \hat{B}_\mathbf{k} + \hat{B}_\mathbf{k}^\dagger \hat{C}_\mathbf{k}),
\end{aligned}
\end{equation}
with
\begin{eqnarray}
\varepsilon^\beta_a(\mathbf{k}) & = & E^\beta_a - 2J^{\beta}_a(\cos k_x \ell + \cos k_y \ell ), \label{eq_disp_ABa} \\
\varepsilon^\beta_B(\mathbf{k}) & = & E^\beta_B - 2J^{\beta}_B(\cos k_x \ell + \cos k_y \ell ) , \\
\varepsilon^{\beta'}_B(\mathbf{k}) & = & - 2W^{\beta}_{B}(\cos k_x \ell - \cos k_y \ell ) , \\
\varepsilon^\beta_d(\mathbf{k}) & = & E^\beta_d - 2J^{\beta}_d(\cos k_x \ell + \cos k_y \ell ). \label{eq_disp_ABd}
\end{eqnarray}
Due to energy degeneracy of the states $\ket{B}$ and $\ket{C}$, there are non-diagonal operators of the form $\hat{C}_\mathbf{k}^\dagger \hat{B}_\mathbf{k}$ and $\hat{B}_\mathbf{k}^\dagger \hat{C}_\mathbf{k}$ within Hamiltonian~\eqref{eq_h_ft}. To remove such terms, we rewrite $\hat{B}_\mathbf{k}$ and $\hat{C}_\mathbf{k}$ in terms of new operators that diagonalise the coupling part, i.e.
\begin{equation} \label{eq_mt_diag}
\begin{aligned}
& \begin{pmatrix}
\hat{B}^\dagger_\mathbf{k} & \hat{C}^\dagger_\mathbf{k}
\end{pmatrix}
\begin{pmatrix}
\varepsilon^\beta_B(\mathbf{k}) & \varepsilon^{\beta'}_B(\mathbf{k}) \\
\varepsilon^{\beta'}_B(\mathbf{k}) & \varepsilon^\beta_B(\mathbf{k})
\end{pmatrix}
\begin{pmatrix}
\hat{B}_\mathbf{k} \\
\hat{C}_\mathbf{k}
\end{pmatrix} \\
& = 
\begin{pmatrix}
\hat{b}^\dagger_\mathbf{k} & \hat{c}^\dagger_\mathbf{k}
\end{pmatrix}
\begin{pmatrix}
\varepsilon^\beta_B(\mathbf{k}) + \varepsilon^{\beta'}_B(\mathbf{k}) & 0 \\
0 & \varepsilon^\beta_B(\mathbf{k}) - \varepsilon^{\beta'}_B(\mathbf{k})
\end{pmatrix}
\begin{pmatrix}
\hat{b}_\mathbf{k} \\
\hat{c}_\mathbf{k}
\end{pmatrix},
\end{aligned}
\end{equation}
where $\hat{b}_\mathbf{k}=\frac{\hat{B}_\mathbf{k}+\hat{C}_\mathbf{k}}{\sqrt{2}}$ and $\hat{c}_\mathbf{k}=\frac{\hat{B}_\mathbf{k}-\hat{C}_\mathbf{k}}{\sqrt{2}}$, allowing Hamiltonian~\eqref{eq_h_ft} to be diagonalised as
\begin{equation}
\begin{aligned}
\hat{H} = & \sum_\mathbf{k} \varepsilon^\beta_a(\mathbf{k}) \hat{a}_\mathbf{k}^\dagger \hat{a}_\mathbf{k} + \sum_\mathbf{k} \varepsilon^\beta_d(\mathbf{k}) \hat{d}_\mathbf{k}^\dagger \hat{d}_\mathbf{k} \\
& + \sum_\mathbf{k} {\varepsilon}^\beta_b(\mathbf{k}) \hat{b}_\mathbf{k}^\dagger \hat{b}_\mathbf{k} + \sum_\mathbf{k} {\varepsilon}^\beta_c(\mathbf{k}) \hat{c}_\mathbf{k}^\dagger \hat{c}_\mathbf{k},
\end{aligned}
\end{equation}
where ${\varepsilon}^\beta_{b,c}(\mathbf{k}) = \varepsilon^\beta_B(\mathbf{k}) \pm \varepsilon^{\beta'}_B(\mathbf{k})$, i.e.
\begin{eqnarray} 
{\varepsilon}^\beta_b(\mathbf{k}) & = & E_B^\beta - 2\big(J^{\beta}_b\cos k_x \ell + J^{\beta}_c\cos k_y \ell \big),
\label{eq_disp_ABb} \\
{\varepsilon}^\beta_c(\mathbf{k}) & = &E_B^\beta - 2\big(J^{\beta}_c\cos k_x \ell + J^{\beta}_b\cos k_y \ell \big),
\label{eq_disp_ABc}
\end{eqnarray}
and
\begin{equation}\label{eq:Js}
J^{\beta}_b = J^{\beta}_B + W^{\beta}_B
\qquad \textrm{and} \qquad
J^{\beta}_c = J^{\beta}_B - W^{\beta}_B.
\end{equation}

From this, we then have $4$ distinct dispersion relations, Eqs.~(\ref{eq_disp_ABa}), (\ref{eq_disp_ABd}), (\ref{eq_disp_ABb}), and (\ref{eq_disp_ABc}), corresponding to four distinct subbands, $\alpha\in\{a,b,c,d\}$.
Subbands $a$ and $d$ have the standard dispersion relation, with different shifts $E_{a,d}^\beta$ and tunnel couplings $J_{a,d}^\beta$.
In contrast the subbands $b$ and $c$ have the same shift but different dispersion relations. Each one, ${\varepsilon}^\beta_b(\mathbf{k})$ and ${\varepsilon}^\beta_c(\mathbf{k})$, are anisotropic and break the $4$-fold rotational symmetry individually. However, the combination of both does not because they are rotated by an angle $\pi/2$ with respect to one another.
In other words, if we rotate the system by $\pi/2$, we recover the same set of dispersion relations.
Similar to band $1$, the energy shifts $E^\beta_{\alpha}$ and band widths $J^\beta_{\alpha}$ can then be extracted by fitting to the continuous dispersion relations, see Sec.~\ref{section_ft}.

\begin{figure}[t!]
	\centering
	\makebox[0pt]{\includegraphics[width=0.99\linewidth]{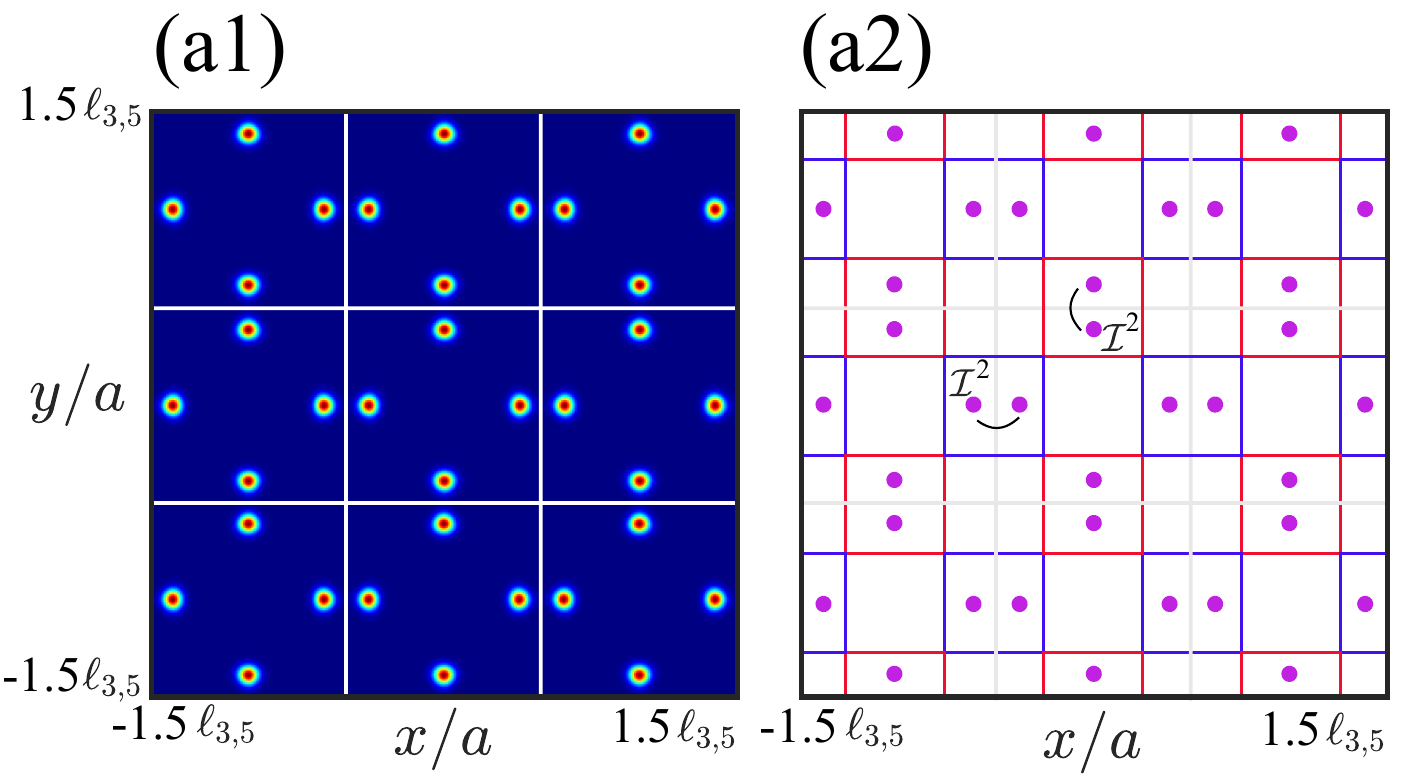}}
	\caption{Plot of the (a1)~density profile for a state in band $2$ and (a2)~the corresponding tight-binding model, taking spots as lattice sites (purple circles). The blue/red squares in (a2) are unit cells of length $\ell/2$, which contain $2$ sites with nearest-neighbour coupling $\mathcal{I}^\beta$ (black). The light grey line in (a2) shows the original moir\'e cell from (a1), with the red/blue cells forming a superlattice structure in this unit cell.}
	\label{fig_band2}
\end{figure}

\subsection{Band $\beta=2$} \label{sec_tbm_b2}
In order to derive the tight-binding model for band $2$, we follow a similar process to bands $3$-$5$, where we separate strong and weak couplings.
For band $2$, however, the sites/spots are now located near the moir\'e unit cell boundaries, so that the dominant couplings are across these boundaries, see Fig.~\ref{fig_band2}(a1).
To derive the relevant tight-binding model, we use a superlattice of strongly-coupled internal states, as depicted in Fig.~\ref{fig_band2}(a2). The smaller blue/red cells, with length $\ell/2$, now contain pairs of sites with the strongest couplings.
This allows us to treat inter-cell couplings in perturbation of intra-cell couplings, consistent with the prior discussions.
Note, however, that we have two types of cells (blue and red), which are identical up to a $\pi/2$-rotation.
Here, the inner Hamiltonian for each blue/red cell is
\begin{equation} \label{eq_h_lt2}
\hat{H}_{cell} = 
\begin{pmatrix}
\epsilon^\beta & -\mathcal{I}^\beta \\
-\mathcal{I}^\beta & \epsilon^\beta
\end{pmatrix},
\end{equation}
which has eigenvalues
\begin{equation}
\begin{aligned}
E^\beta_{A} & = \epsilon^\beta + \mathcal{I}^\beta, \\ E^\beta_{B} & = \epsilon^\beta - \mathcal{I}^\beta,
\end{aligned}
\end{equation}
and normalised eigenstates
\begin{equation} \label{eq_eigVec2}
\begin{aligned}
& \ket{A} = \ket{A'} = \frac{1}{\sqrt{2}}
\begin{pmatrix}
1 \\ 1 
\end{pmatrix},
\\
& \ket{B} = \ket{B'} = \frac{1}{\sqrt{2}}
\begin{pmatrix}
1 \\ -1
\end{pmatrix},
\end{aligned}
\end{equation}
where the eigenstates ($\ket{A}$, $\ket{B}$) and ($\ket{A'}$, $\ket{B'}$) belong to the blue and red cells in Fig.~\ref{fig_band2}(a2) respectively.
Note, the implicit basis used to write the Hamiltonian and the eigenstates, Eqs.~(\ref{eq_h_lt2}) and (\ref{eq_eigVec2}), is rotated with an angle $\pi/2$ for red cells with respect to blue cells.
This creates two energy-separated subbands, $A-A'$ on the one hand and $B-B'$ on the other hand, with negligible couplings between the two.
In contrast, couplings between $A$ and $A'$ states, which have degenerate on-site energies, must be taken into account.
\begin{figure}[t!]
	\centering
	\makebox[0pt]{\includegraphics[width=0.99\linewidth]{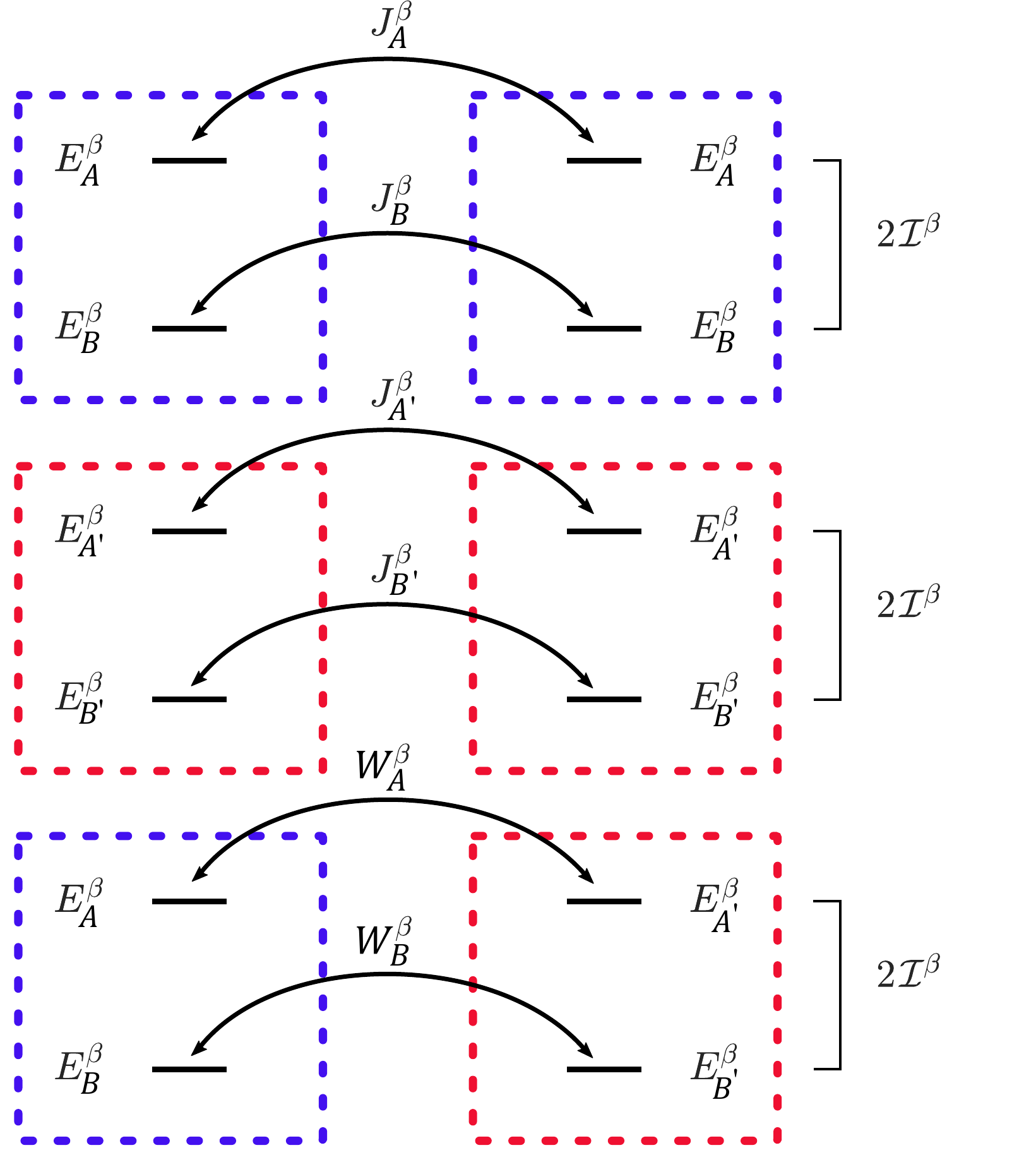}}
	\caption{Coupling of states from Eq.~\eqref{eq_h_lt2} between red-to-red, blue-to-blue and red-to-blue cells of Fig.~\ref{fig_band2}(a2). Black lines denote the states of each subband $\alpha$, with corresponding energy $E_\alpha^\beta$. Due to $4$-fold rotational symmetry, we have $E_A^\beta=E_{A'}^\beta$ and $E_B^\beta=E_{B'}^\beta$. The energy spacing between states is shown outside the righthand cells.}
	\label{fig_band2_coupling}
\end{figure}
The same holds for $B$ and $B'$ states.
By coupling the different cells as per Fig.~\ref{fig_band2_coupling} and retaining only the nearest-neighbour terms of each type, the Hamiltonian for band $\beta=2$ can then be expressed as
\begin{equation} \label{eq_band2H}
\hat{H} = \hat{H}_A + \hat{H}_B,
\end{equation}
with
\begin{eqnarray}
\hat{H}_A
& = & E^\beta_{A} \sum_{i} (\hat{A}_i^\dagger \hat{A}_i + \hat{A}_i^{'\dagger} \hat{A}^{'}_i)
\nonumber \\
&& - \sum_{\langle i,j \rangle \in R\leftrightarrow B} W^{\beta}_A \hat{A}_i^\dagger \hat{A}^{'}_j \\
&& - \sum_{\langle i,j \rangle \in B\rightarrow B} J^{\beta}_{A;i,j}\hat{A}_i^\dagger \hat{A}_j
      - \sum_{\langle i,j \rangle \in R\rightarrow R} J^{\beta}_{A';i,j}\hat{A}_i^{'\dagger} \hat{A}^{'}_j
\nonumber
\end{eqnarray}
and
\begin{eqnarray}
\hat{H}_B
& = & E^\beta_{B} \sum_{i} (\hat{B}_i^\dagger \hat{B}_i + \hat{B}_i^{'\dagger} \hat{B}^{'}_i)
\nonumber \\
&& - \sum_{\langle i,j \rangle \in B\leftrightarrow R} W^{\beta}_B \hat{B}_i^\dagger \hat{B}^{'}_j \\
&& - \sum_{\langle i,j \rangle \in B\rightarrow B} J^{\beta}_{B;i,j}\hat{B}_i^\dagger \hat{B}_j
      - \sum_{\langle i,j \rangle \in R\rightarrow R} J^{\beta}_{B';i,j}\hat{B}_i^{'\dagger} \hat{B}^{'}_j,
\nonumber
\end{eqnarray}
where $\langle i,j \rangle \in B\leftrightarrow R$ denotes a sum over pairs of nearest neighbouring cells from both a blue cell to a red cell and a red cell to a blue cell,
$\langle i,j \rangle \in B\rightarrow B$ between blue cells,
and $\langle i,j \rangle \in R\rightarrow R$ between red cells.
Since the red cells (corresponding to the primed symbols) are obtained from a $\pi/2$-rotation of blue cells (corresponding to the non-primed symbols),
we must have $J^{\beta}_{A;x}=J^{\beta}_{A';y}$, $J^{\beta}_{A;y}=J^{\beta}_{A';x}$, $J^{\beta}_{B;x}=J^{\beta}_{B';y}$, and $J^{\beta}_{B;y}=J^{\beta}_{B';x}$, see Appendix~\ref{app_latticeIV}.
In the following, we restrict the discussion to the subband $A-A'$. Since the Hamiltonians $\hat{H}_A$ and $\hat{H}_B$ have the same structure, all formulas for the subband $B-B'$ are the same as for the subband $A-A'$ replacing $A$'s by $B$'s.

By transforming the operators to momentum space, the Hamiltonian reads as
\begin{eqnarray} \label{eq_h_ft2}
\hat{H}_A & = & \sum_\mathbf{k} \left( \varepsilon^\beta_{AA}(\mathbf{k})\hat{A}_\mathbf{k}^\dagger \hat{A}_\mathbf{k} + \varepsilon^\beta_{A'A'}(\mathbf{k})\hat{A}^{'\dagger}_\mathbf{k} \hat{A}^{'}_\mathbf{k} \right) \nonumber \\
&& + \sum_\mathbf{k} \varepsilon^\beta_{AA'}(\mathbf{k})( \hat{A}_\mathbf{k}^\dagger \hat{A}^{'}_\mathbf{k} + \hat{A}^{'\dagger}_\mathbf{k} \hat{A}_\mathbf{k}).
\end{eqnarray}
where
\begin{equation}\label{eq:band2epsilons}
\begin{aligned}
\varepsilon^\beta_{AA}(\mathbf{k}) & = E^\beta_{A} - 2(J^{\beta}_{A;x}\cos k_x \ell + J^{\beta}_{A;y}\cos k_y \ell ), \\
\varepsilon^\beta_{A'A'}(\mathbf{k}) & = E^\beta_{A} - 2(J^{\beta}_{A;y}\cos k_x \ell + J^{\beta}_{A;x}\cos k_y \ell ), \\
\varepsilon^\beta_{AA'}(\mathbf{k}) & = -4W^{\beta}_A\Big(\cos \frac{k_x \ell}{2} \cos \frac{k_y \ell}{2} \Big).
\end{aligned}
\end{equation}
To diagonalise the problem, we then use a similar procedure to before, where we introduce new operators $\hat{a}$, $\hat{a}^{'}$, $\hat{a}^{\dagger}$, and $\hat{a}^{'\dagger}$, which diagonalise terms involving $\hat{A}$ and $\hat{A}^{'}$,  $\hat{A}^{\dagger}$, and $\hat{A}^{'\dagger}$,
\begin{equation} \label{eq_mt_diag2}
\begin{aligned}
\begin{pmatrix}
\hat{A}^\dagger_\mathbf{k} & \hat{A}^{'\dagger}_\mathbf{k}
\end{pmatrix}
\begin{pmatrix}
\varepsilon^\beta_{AA}(\mathbf{k}) & \varepsilon^{\beta'}_{AA'}(\mathbf{k}) \\
\varepsilon^{\beta'}_{AA'}(\mathbf{k}) & \varepsilon^\beta_{A'A'}(\mathbf{k})
\end{pmatrix}
\begin{pmatrix}
\hat{A}_\mathbf{k} \\
\hat{A}^{'}_\mathbf{k}
\end{pmatrix} \\
=
\begin{pmatrix}
\hat{a}^\dagger_\mathbf{k} & \hat{a}^{'\dagger}_\mathbf{k}
\end{pmatrix}
\begin{pmatrix}
\varepsilon^\beta_{a}(\mathbf{k}) & 0 \\
0 & \varepsilon^\beta_{a'}(\mathbf{k})
\end{pmatrix}
\begin{pmatrix}
\hat{a}_\mathbf{k} \\
\hat{a}^{'}_\mathbf{k}
\end{pmatrix},
\end{aligned}
\end{equation}
where
\begin{equation}\label{eq:band2a}
\begin{aligned}
& {\varepsilon}^\beta_{a}(\mathbf{k}) = \\
& \frac{\varepsilon^\beta_{AA}(\mathbf{k}) \! + \! \varepsilon^\beta_{A'A'}(\mathbf{k}) \! + \! \sqrt{\left(\varepsilon^\beta_{AA}(\mathbf{k}) \! - \!\varepsilon^\beta_{A'A'}(\mathbf{k})\right)^2 \!+ \!4\varepsilon^{\beta}_{AA'}(\mathbf{k})^2 }}{2}
\\
\end{aligned}
\end{equation}
and
\begin{equation}\label{eq:band2ap}
\begin{aligned}
& {\varepsilon}^\beta_{a'}(\mathbf{k}) = \\
& \frac{\varepsilon^\beta_{AA}(\mathbf{k}) \!+ \!\varepsilon^\beta_{A'A'}(\mathbf{k}) \!- \!\sqrt{\left(\varepsilon^\beta_{AA}(\mathbf{k})- \!\varepsilon^\beta_{A'A'}(\mathbf{k})\right)^2 \!+ \!4\varepsilon^{\beta}_{AA'}(\mathbf{k})^2}}{2},\\
\end{aligned}
\end{equation}
and similar formulas for the $B-B'$ subband. 
The final Hamiltonian is then
\begin{equation}
\begin{aligned}
\hat{H} = \sum_\mathbf{k} \varepsilon^\beta_{{a}}(\mathbf{k}) \hat{{a}}_\mathbf{k}^\dagger \hat{{a}}_\mathbf{k} + \sum_\mathbf{k} \varepsilon^\beta_{{a}^{'}}(\mathbf{k}) \hat{{a}}^{'\dagger}_\mathbf{k} \hat{{a}}^{'}_\mathbf{k}  \\ + \sum_\mathbf{k} \varepsilon^\beta_{{b}}(\mathbf{k}) \hat{{b}}_\mathbf{k}^\dagger \hat{{b}}_\mathbf{k} + \sum_\mathbf{k} \varepsilon^\beta_{{b}^{'}}(\mathbf{k}) \hat{{b}}^{'\dagger}_\mathbf{k} \hat{{b}}^{'}_\mathbf{k}.
\end{aligned}
\end{equation}
We thus find four distinct subbands $\alpha\in\{a,a',b,b'\}$ within band $\beta=2$.
We can then extract the energy shifts $E_A^{\beta}$ and $E_B^{\beta}$, as well as the inter-cell tunnel energies $J_{A;x}^{\beta}$, $J_{A;y}^{\beta}$, $J_{A;x}^{\beta'}$, $J_{B;x}^{\beta}$, $J_{B;y}^{\beta}$, and $J_{B;x}^{\beta'}$ by fitting the continuous-space dispersion relations, see Appendix~\ref{app:tbm_val_c4}.
Note, these bands have non-standard dispersion relations, given by Eqs.~(\ref{eq:band2a}) and (\ref{eq:band2ap}), with Eq.~(\ref{eq:band2epsilons}). In most cases, couplings between blue-to-red cells dominate over those between cells with the same colour, due to the smaller distance, see Fig.~\ref{fig_band2}(a2).
We then have
\begin{equation}\label{eq:band2aapApprox}
\begin{aligned}
& {\varepsilon}^\beta_{a,a'}(\mathbf{k}) \approx E_A^\beta \pm \varepsilon^\beta_{AA'}(\mathbf{k}) = E_A^\beta \mp 4W^{\beta}_A\Big(\cos \frac{k_x \ell}{2} \cos \frac{k_y \ell}{2} \Big),
\\\\
& {\varepsilon}^\beta_{b',b'}(\mathbf{k}) \approx E_B^\beta \pm \varepsilon^\beta_{BB'}(\mathbf{k}) = E_B^\beta \mp 4W^{\beta}_B\Big(\cos \frac{k_x \ell}{2} \cos \frac{k_y \ell}{2} \Big),
\\
\end{aligned}
\end{equation}
which will be illustrated in Sec.~\ref{section_ft}. Note, the unit cell with only blue-to-red couplings has a smaller square length of $\ell/\sqrt{2}$, and larger square Brillouin zone length of $2\sqrt{2}\pi/\ell$. The momentum dependence can be written as
\begin{equation}
\begin{aligned}
&4W^{\beta} \cos \frac{k_x \ell}{2} \cos \frac{k_y \ell}{2} = 2W^{\beta} \left( \cos \frac{\bar{k}_x \ell}{\sqrt{2}} + \cos \frac{\bar{k}_y \ell}{\sqrt{2}} \right),
\end{aligned}
\end{equation}
with $\bar{k}_x = \mathbf{k} \cdot \bar{\mathbf{x}}$ and $\bar{k}_y = \mathbf{k} \cdot \bar{\mathbf{y}}$, where $\bar{\mathbf{x}}=(\hat{\mathbf{x}}+\hat{\mathbf{y}})/\sqrt{2}$ and $\bar{\mathbf{y}}=(\hat{\mathbf{x}}-\hat{\mathbf{y}})/\sqrt{2}$ are unit vectors of the smaller (rotated) square lattice. The subbands $a$ and $a'$ are thus centred around energy $E_A^\beta$ and have almost opposite variations with $\mathbf{k}$.
The same holds for subbands $b$ and $b'$, but they are centred around a different energy, $E_B^\beta$, and the amplitude of variations is different from that of subbands $a$ and $a'$.

\begin{figure}[t!]
	\centering
	\makebox[0pt]{\includegraphics[width=0.99\linewidth]{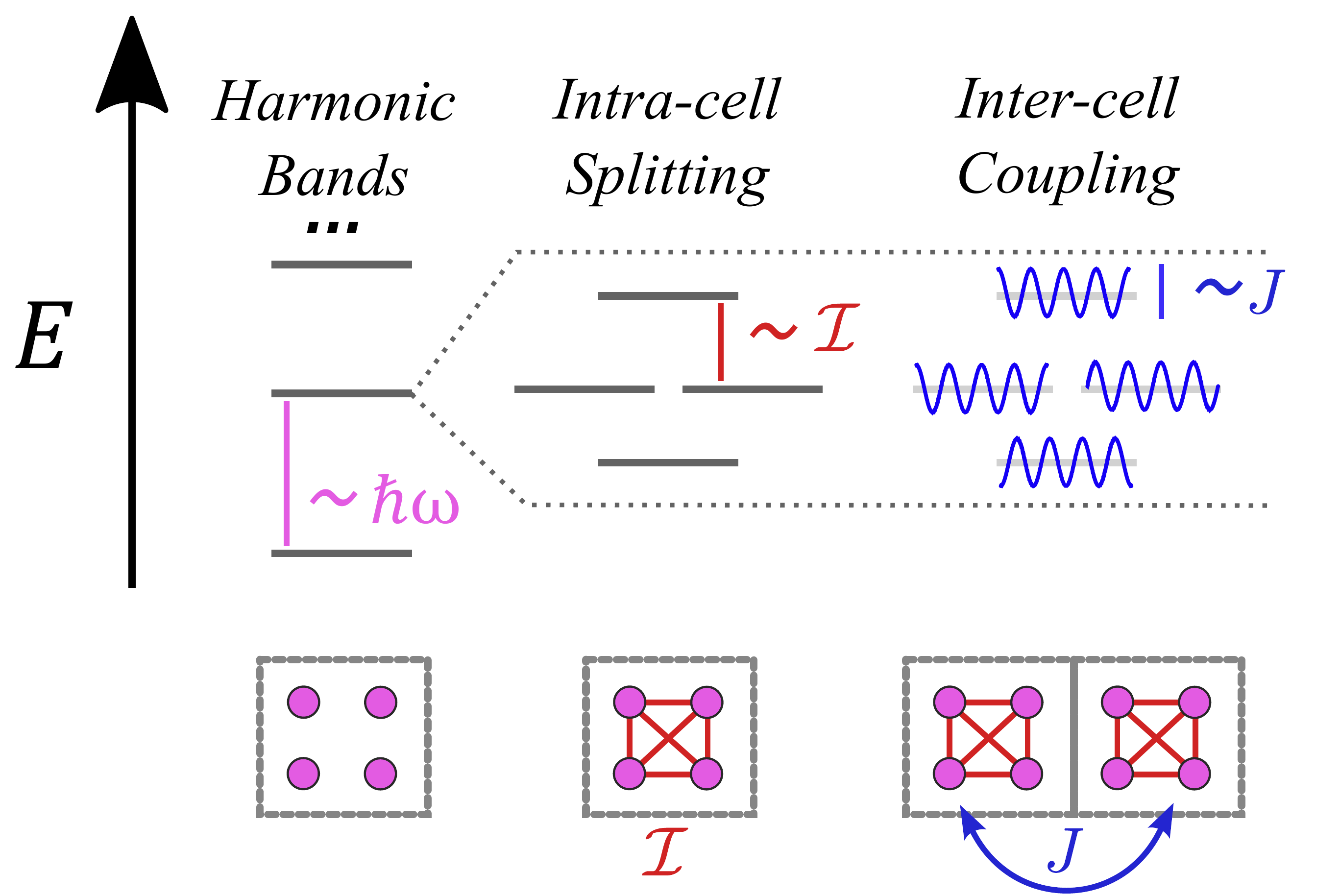}}
	\caption{Energy scales within twisted moir\'e potentials. On the left, we have degenerate harmonic bands with characteristic separation $\hbar \omega$, corresponding to the $4$ sites (purple circles in the lower grey square) of an isolated moir\'e cell. By coupling the sites from a given unit cell, degenerate harmonic bands are split into subbands, with separation depending on the intra-cell tunnelling $\mathcal{I}$. Finally, by coupling nearest-neighbour moir\'e cells, the energy spectrum will exhibit cosine-like dispersions for each subband, with width scaled to the inter-cell tunnelling $J$.}
	\label{fig_modelEn}
\end{figure}

\begin{figure*}[t!]
	\centering
	\makebox[0pt]{\includegraphics[width=0.99\linewidth]{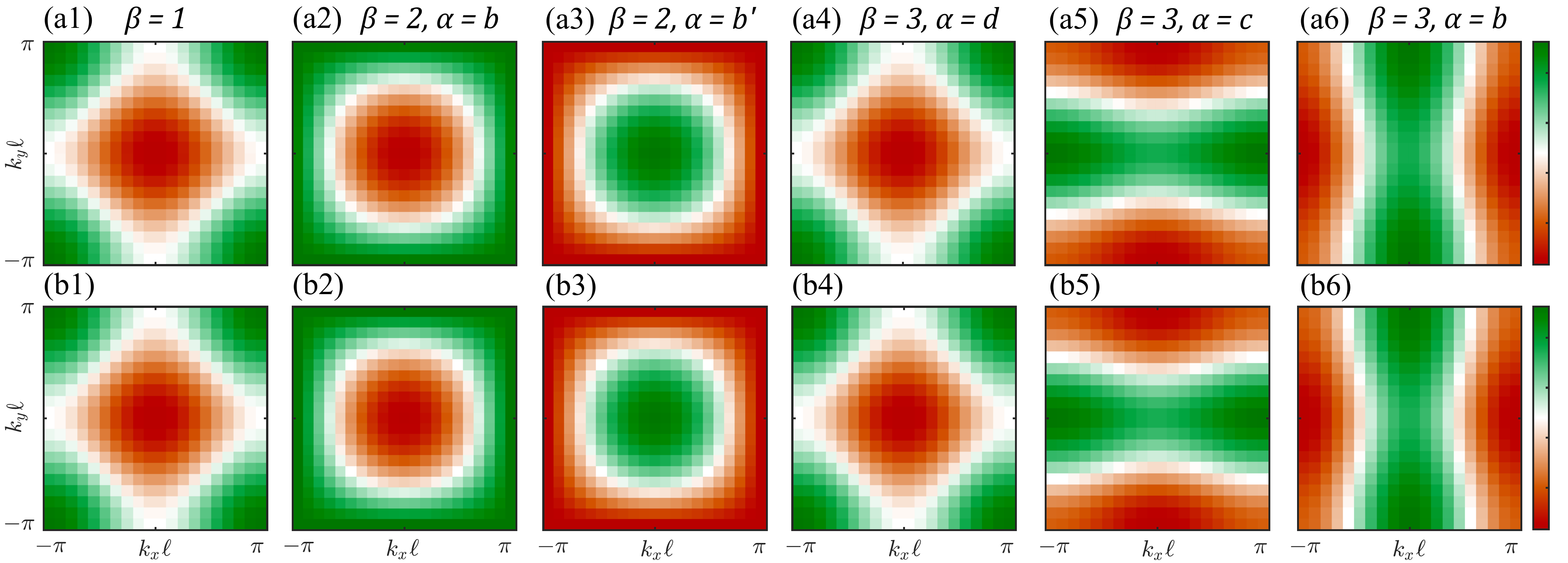}}
	\caption{Momentum ($\mathbf{k}$) dependence of the dispersion relations from the (a1)-(a6) continuous results when $V=6\Er$ and (b1)-(b6) fitted tight-binding models. The colourscales only show the momentum dependence of each dispersion relation, i.e. without an onsite energy offset and width scaling by tunnelling rates. We consider bands and subbands (a1),(b1) $\beta=1$; (a2),(b2) $\beta=2$ and $\alpha=b$; (a3),(b3) $\beta=2$ and $\alpha=b'$; (a4),(b4) $\beta=3$ and $\alpha=d$; (a5),(b5) $\beta=3$ and $\alpha=c$; (a6),(b6) $\beta=3$ and $\alpha=b$. In all cases, lattice and continuous results are in good agreement. Equivalent properties are also observed for the other bands and subbands.}
	\label{fig_bands_disp2D_NEW}
\end{figure*}

\section{Effective Tight-Binding Parameters} \label{section_ft}

So far, we have separated different energy scales in order to characterise the full, continuous spectra. In Fig.~\ref{fig_modelEn}, we illustrate the different energy scales present in twisted moir\'e potentials as a reminder. First, we have the basic structure of well-separated bands, which are captured with a harmonic approximation around each minima. Next, by coupling different minima in a moir\'e cell together, degenerate harmonic states are split into distinct subbands, with intra-cell tunnellings governing the separation between each subband. Finally, by coupling different moir\'e cells together, we introduce the final energy scale of inter-cell couplings, which produces the cosine-like features in the dispersion relations.

We now compare the predictions of the effective tight-binding models constructed in Sec.~\ref{section_tbm} with the exact results obtained in Sec.~\ref{section_spect} from the continuous-space model. The parameters of tight-binding models are obtained by fitting the tight-binding dispersion relation for each subband to the corresponding one for the continuous-space model.
We show the momentum dependence of some example bands and subbands in Fig.~\ref{fig_bands_disp2D_NEW}, comparing continuous (upper row) and tight-binding (lower row) dispersion relations at the moir\'e angle of $\theta_{3,5}$. These plots show the momentum dependencies only, i.e. onsite energy offsets are set to zero and modulations of cosine functions are set to unity. From the comparisons, we immediately see that the momentum dependence of the continuous-space dispersions is accurately captured with the tight-binding models. For the ground state subband $\beta=1$ in the first column of Fig.~\ref{fig_bands_disp2D_NEW}, the dispersion relation follows the standard dispersion of a 2D square lattice, Eq.~(\ref{eq_k_sq}).
In the second and third columns, corresponding to $\beta=2$, we show two quasi-degenerate subbands $\alpha=(b,\,b')$. The dispersion relations are consistent with the non-standard forms given in Eqs.~(\ref{eq:band2a}) and (\ref{eq:band2ap}), with Eq.~(\ref{eq:band2epsilons}). We also find good agreement with the approximation of Eq.~(\ref{eq:band2aapApprox}). Next, in band $\beta=3$, corresponding to the last $3$ columns in Fig.~\ref{fig_bands_disp2D_NEW}, we have two non-degenerate subbands $\alpha=d$ (fourth column) and $\alpha=a$ (not shown), with dispersions again following that of a standard 2D square lattice. Finally, in the last two columns, we show the degenerate subbands $\alpha=c$ and $\alpha=b$, which have strongly anisotropic dispersions, arising due to the $\pi/2$-rotational symmetry as discussed in the prior section, consistent with Eqs.~(\ref{eq_disp_ABb}) and (\ref{eq_disp_ABc}).

More precisely, we quantify the agreement between the continuous-space and tight-binding models with a residual parameter $\gamma^\beta_\alpha$, defined as the average difference between the tight-binding $\varepsilon_\textrm{TBM}$ and continuous-space $\varepsilon_\textrm{C}$ dispersion relations,
\begin{equation} \label{eq_gamma}
\begin{aligned}
\gamma^\beta_\alpha = \frac{\ell^2}{4\pi^2} \int_{-\pi/\ell}^{\pi/\ell} \int_{-\pi/\ell}^{\pi/\ell} \, dk_x \, dk_y \, \left|\varepsilon_\textrm{TBM} (\mathbf{k}) - \varepsilon_\textrm{C} (\mathbf{k})\right|.
\end{aligned}
\end{equation}
Since the dispersions are scaled
by the tunnelling rates $J^\beta_\alpha$, we consider the ratio $\gamma^\beta_\alpha / J^\beta_\alpha$ for each band and subband in order to provide meaningful comparisons. For a tight-binding model to be accurate, we typically require that $\gamma^\beta_\alpha / J^\beta_\alpha \ll 1$. In Fig.~\ref{fig_bands_disp2D_NEW}, we find that the largest residuals are $\gamma^\beta_\alpha / J^\beta_\alpha \sim 0.04$, i.e.~no noticeable differences between dispersion relations. These errors may be further reduced by including beyond nearest-neighbour couplings within the tight-binding models. However, for this work, we find it sufficient to only consider couplings with separation up to the moir\'e length $\ell$ for high accuracy across a range of potential depths.

\begin{figure*}[t!]
	\centering
	\makebox[0pt]{\includegraphics[width=0.8\linewidth]{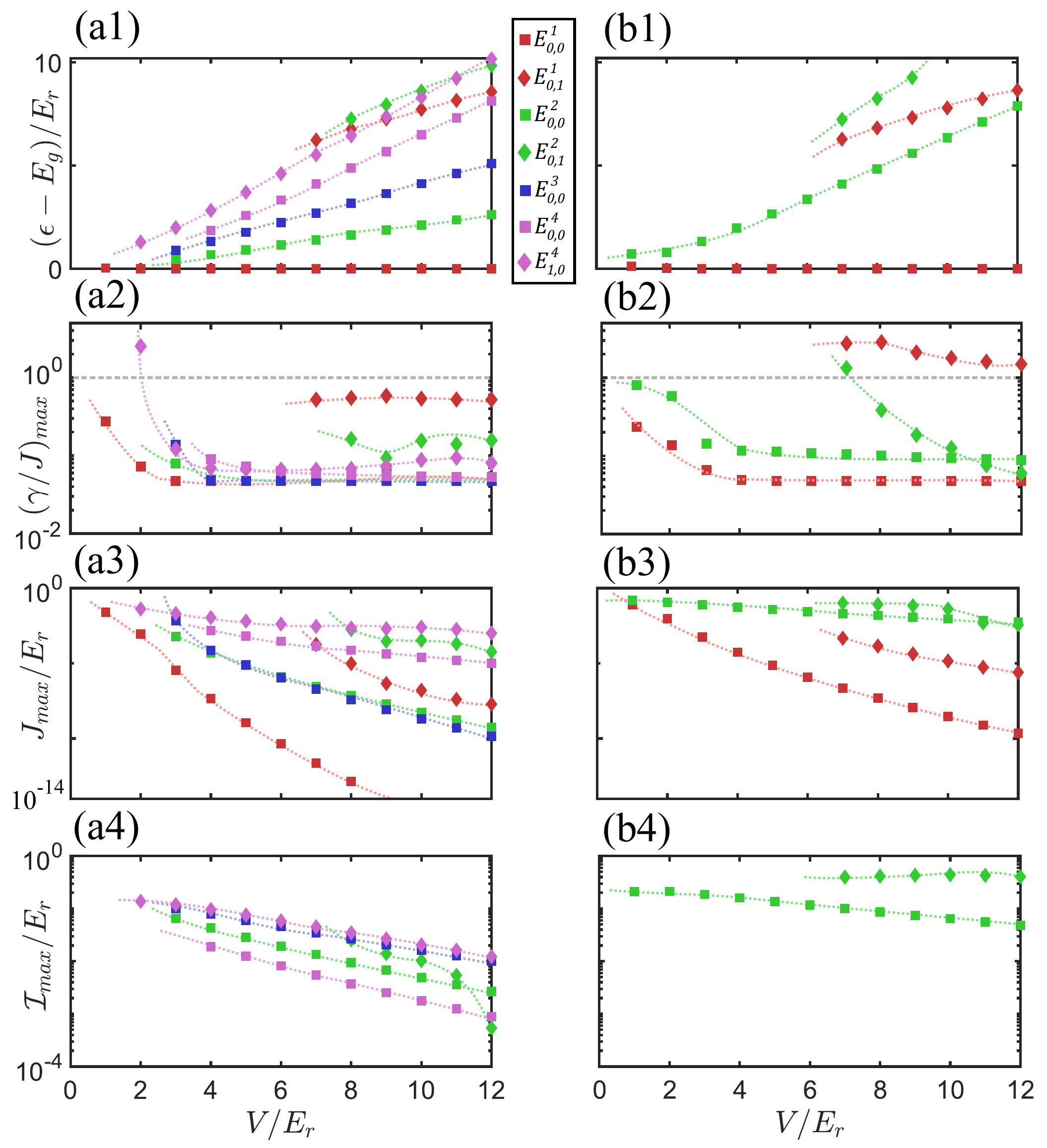}}
	\caption{Tight-binding parameters as a function of $V/\Er$, for angles (a1)-(a4) $\theta_{3,5}$ and (b1)-(b4) $\theta_{2,1}$. For different bands, we plot the (a1),(b1) onsite energy $\epsilon$, the (a2),(b2) largest residual $(\gamma/J)_{max}$, (a3),(b3) largest intercell coupling $J_{max}$ and the (a4),(b4) largest intracell coupling $\mathcal{I}_{max}$. Different bands are labelled according to the related harmonic energy $E^u_{n_+,n_-}$ (see main text), with colours for the $u$-th minima as in Fig.~\ref{fig_harmonicC_min}, where squares are ground states and diamonds are excited states. Coloured dashed lines are guides to the eye, with the grey line in (a2),(b2) showing $(\gamma/J)_{max}=1$, i.e.~the effective threshold for the tight-binding validity. }
	\label{fig_varyV}
\end{figure*}

For smaller potential depths $V/\Er$, it is to be expected that the effective tunnelling rates of tight-binding models will increase, but at the same time the validity of the tight-binding approximation should progressively deteriorate.
To understand the scaling of tight-binding parameters and the range of validity, we consider the moir\'e potential for a range of potential depths $V$ 
and two twist angles, $\theta_{3,5}$ in Figs.~\ref{fig_varyV}(a1)-(a4) and  $\theta_{2,1}$ in Figs.~\ref{fig_varyV}(b1)-(b4). Here, different bands are coloured according to the $u$-th minima in Fig.~\ref{fig_harmonicC_min} and labelled according to the equivalent harmonic energy/state $E^u_{n_+,n_-}$. The matching between states is performed by comparing the structure and localisation of a continuous state with harmonic eigenstates, e.g.~continuous states that are Gaussian-like around some minima $u$ are labelled as $E^u_{0,0}$ and likewise for higher excitations in $n_\pm$. As discussed, different minima corresponds to different ladders of energy bands, which may cross at certain values of $V/\Er$. We then plot some of the lowest ground state and excited bands to illustrate this. In particular, we plot $\epsilon=\epsilon^\beta$, the intra-cell, or onsite energy of band $\beta$, which is equivalent to the average shift of subband energy, see Appendix~\ref{app:tbm_val}. We also plot $(\gamma/J)_{max}$, which is defined as the largest $\gamma^\beta_\alpha / J^\beta_\alpha$ for band $\beta$. Finally, we also plot $J_{max}$ (the largest intercell tunnelling of a band, in magnitude) and $\mathcal{I}_{max}$ (the largest intracell tunnelling of a band, in magnitude). Note, we take the absolute values of tunnellings to determine the maximum, since both positive and negative tunnellings exist within different bands. In Figs.~\ref{fig_varyV}(a1),(b1), we observe a familiar structure of energy gaps in the spectrum to that of Fig.~\ref{fig_harmonicC}, with crossings at larger $V\simeq 9\Er$ (red and purple diamonds) and $V\simeq 11\Er$ (green and purple diamonds) for the excited harmonic states.
We find that the energy shifts increase with the potential depth. This is to be expected, since they are governed by the energies at the potential minima (proportional to $V$) and the frequencies of the harmonic approximation (proportional to $\sqrt{V}$).
The quasi-linear behaviour observed here suggests that the energy shifts are dominated by the former contribution while the latter is negligible.
To show that the tight-binding models are indeed valid across a range of $V$, we plot the largest residuals $(\gamma/J)_{max}$ in Figs.~\ref{fig_varyV}(a2),(b2), which shows that $(\gamma/J)_{max} \ll 1$ for the majority of bands.
For bands generated from excited harmonic states (diamonds), the accuracy of the tight-binding model is not as good for the considered potential amplitudes, owing to larger on-site energies. We, however, expect to find good agreement for larger potential amplitude, consistently with the observed tendency of decreasing $\gamma/J$ with $V$.
The inter-cell ($J_{max}$) and intra-cell ($\mathcal{I}_{max}$) couplings decrease exponentially
with $V$, as observed in Figs.~\ref{fig_varyV}(a3)-(b3) and (a4)-(b4).
This is also to be expected since the potential barriers increase with $V$.

When considering large $V/\Er$, all tunnellings and $\gamma^\beta_\alpha$ become smaller. In other words, each potential minima/site of the tight-binding models are becoming decoupled, and can be described more accurately via the harmonic approximation introduced in Sec.~\ref{section_spect}. As $V/\Er$ becomes smaller, coupling between potential minima/sites lift the degeneracies that are associated to the harmonic approximation. However, localised Wannier functions can still form in the low energy bands, resulting in very small values of $(\gamma/J)_{max}$, i.e.~tight-binding theory can still be applied. Generally speaking, higher energy bands will have larger values of $(\gamma/J)_{max}$, due to the more extended behaviour of the associated Wannier functions across space. While visible distortions to the full 2D dispersion are small, the continuous dispersion relation is no longer modulated by nearest-neighbour tunnelling alone, meaning that $(\gamma/J)_{max}$ is more sensitive to small fluctuations. Finally, it is also important to note that at smaller $V/\Er$, the subbands of certain bands will no longer be isolated in energy from other bands, hence the considered tight-binding framework can not be applied. For $\theta_{3,5}$, this occurs for the $u=2,3,4$ minima around $V/\Er\sim3-4$. Similar properties also occur for excited states within different minima, e.g. $E^1_{0,1}$ and $E^2_{0,1}$ bands being absent for $V/\Er \lesssim 7$ (red/green diamonds).

The inter-cell tunnellings in Fig.~\ref{fig_varyV}(a3) usually grow in magnitude as the band index increases, giving the approximate ordering $J^1_\alpha < J^2_\alpha < J^3_\alpha \dots$ and likewise for the intra-cell couplings in Fig.~\ref{fig_varyV}(a4). The reason behind this can be linked to the fact that higher band indices denote states that are either localised in higher energy potential minima, or excited Wannier states in the potential minima, which extend further away from the minima and thus enhance tunnelling probabilities. Note, there are some special cases where this may not occur, e.g.~between bands $2$ and $3$ in Fig.~\ref{fig_varyV}(a3) (green and blue squares), where the nearest-neighbour inter-cell couplings have the separation $\sim \ell/\sqrt{2}$ rather than $\sim \ell$ due to the superlattice structure. For more general or more exotic moir\'e potentials, a similar breakdown may also occur.

When considering different moir\'e angles, the distribution of tight-binding parameters can drastically change. In Figs.~\ref{fig_varyV}(b1)-(b4), we plot the same results for the commensurate angle $\theta_{2,1}$. For this angle, we have two distinct minima as per Fig.~\ref{fig_harmonicC_min}(b), where bands and subbands can be modelled using the same tight-binding models as before, with red points equivalent to band $1$ from Sec.~\ref{sec_tbm_b1} and green points equivalent to band $4$ from Sec.~\ref{sec_tbm_b345}.
The energy separation between minima for the commensurate angle $\theta_{2,1}$ is larger, allowing for the formation of more prominent spectral gaps in Fig.~\ref{fig_varyV}(b1). Furthermore, the spatial separation between minima is smaller, thus enhancing the observable tunnelling rates in Figs.~\ref{fig_varyV}(b3)-(b4). This contrasts the behaviour observed for $\theta_{3,5}$ from Figs.~\ref{fig_varyV}(a1)-(a4), in which the moir\'e cell is enlarged, i.e.~we have more distinct potential minima, with smaller relative separations in energy. This leads to more, but smaller gaps and tunnellings, since the minima are distributed further across space.

\begin{figure*}[t!]
	\centering
	\makebox[0pt]{\includegraphics[width=0.99\linewidth]{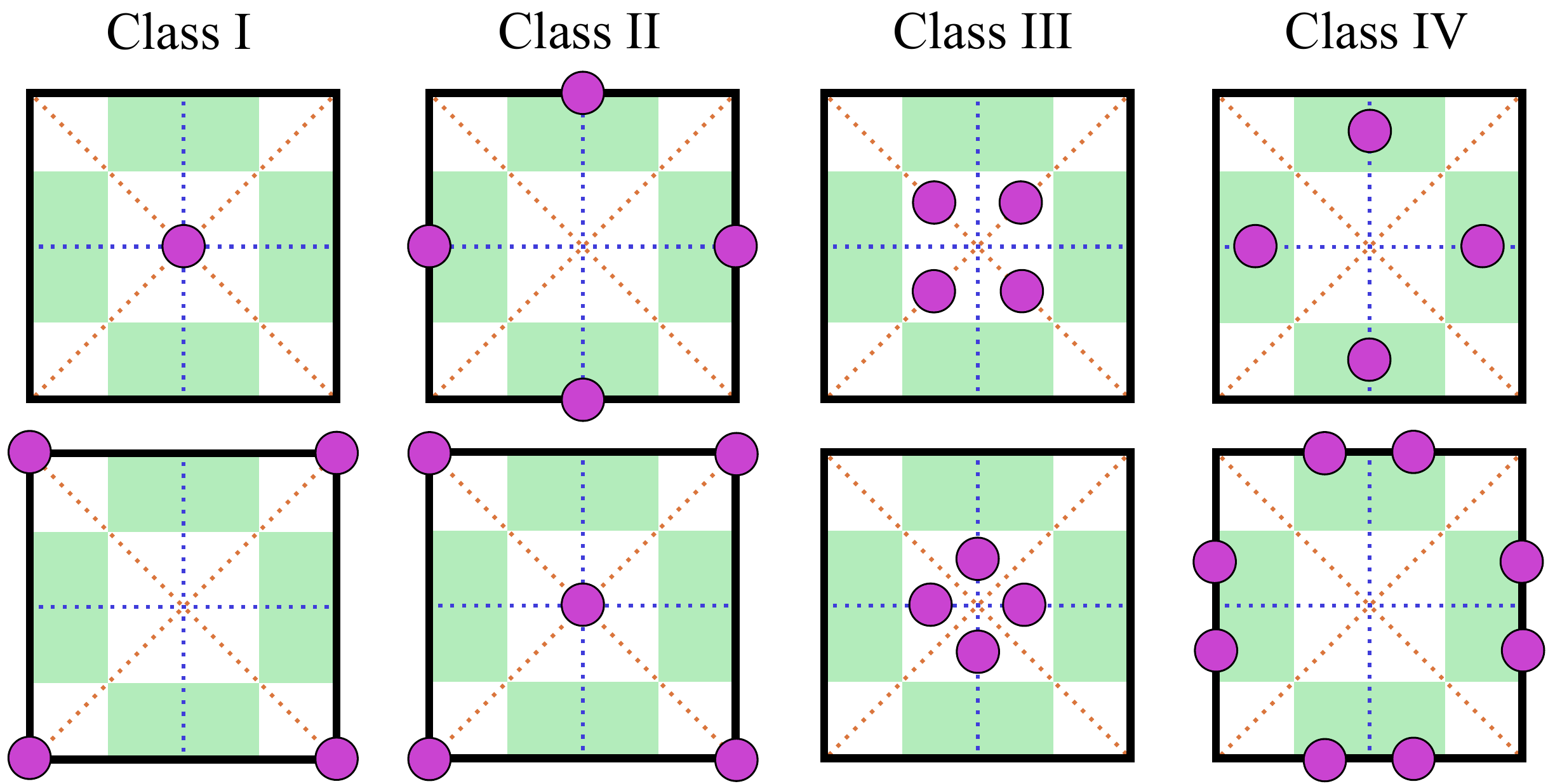}}
	\caption{Unique arrangements of potential minima in a moir\'e cell with $4$-fold rotational symmetry (blue/orange dotted lines are symmetry lines), ordered into classes of tight-binding models/bands. Green areas denote regions of space that will always be closer to an adjacent moir\'e cell than the centre, regardless of a shift to the cell. For Class I bands, we have one minima in the centre of the cell, or, by shifting the cell, $1/4$ minima at the $4$ corners. For Class II bands, we have $1/2$ minima at the midpoints of each edge, or, by shifting the cell, $1$ minima in the centre with $1/4$ minima at $4$ corners. Class III bands have $4$ minima in the cell along symmetry lines, outside the green regions. Finally, Class IV bands have $4$ minima inside the green regions, or, by shifting the unit cell, $8 \, 1/2$ minima along the edges.}
	\label{fig_bandClasses}
\end{figure*}

\section{Generalisation: Classification of Bands}\label{section_general}
So far, the bands we have studied for $\theta=\theta_{3,5}$ and  $\theta=\theta_{2,1}$ fall into $3$ distinct classes. For completeness, we now discuss the geometrical classification of bands for arbitrary $\theta$. When $V/\Er$ is deep enough,
the low energy eigenstates are well-localised around distinct sets of potential minima.
Each set
forms a distinct geometry for an underlying tight-binding model, with examples from the prior sections.

Based on the $4$-fold rotational symmetry and the fact that there are at most $4$ degenerate minima in the moir\'e cell, we may generally identify $4$ unique classes of bands or geometrical arrangements that can appear in twisted square moir\'e potentials.
These are shown in Fig.~\ref{fig_bandClasses}, with the green areas denoting regions of space where a site is closer to the moir\'e cell boundary than the centre (regardless of a shift to the cell).
In practice, we find that the shortest distance between sites determines the strongest couplings. It may, however, not be excluded that some short-distance couplings are suppressed by high potential barriers. Hence, more precisely, in our classification, sites in the white regions correspond to cases where intra-cell couplings dominate over inter-cell couplings (as for bands $1$, $3$, $4$, and $5$ discussed above), while, on the contrary, sites in the green regions correspond to cases where inter-cell couplings dominate over intra-cell couplings (as for band $2$ discussed above).
Note that the cases on the upper and lower rows of the same column in Fig.~\ref{fig_bandClasses} are actually equivalent upon a shift or rotation of the unit moir\'e cell.

\subsection{Class I}
We start with the simplest case of Class I in Fig.~\ref{fig_bandClasses}. Here, we have effectively $1$ potential minima enclosed by the moir\'e cell, with components located at the cell center or corners. The dispersion is then that of a square lattice, which was discussed in Sec.~\ref{sec_tbm_b1}. Sites of Class I bands are always located at the global minima of the potential. There is one subtle point to note about these bands, however. From the harmonic spectrum at larger $V/\Er$ in Fig.~\ref{fig_harmonicC}(a1,b1), the $n$-th excited states in these minima are implied to be $(n+1)$-fold degenerate, which would require a generalised matrix diagonalisation procedure to describe the bands. However, in the continuous spectrum, anharmonic terms will lift this degeneracy, well beyond that of the tunnelling rates, as can be seen more clearly in Fig.~\ref{fig_harmonicC}(b1), with the thicker grey lines denoting non-degenerate states. In other words, excited states that fall under the Class I specification can also be described by the model in Sec.~\ref{sec_tbm_b1}.

\subsection{Class II}
In Fig.~\ref{fig_bandClasses}, we also show the configuration of Class~II bands. Here, we have $4$ potential minima at the midpoints of the unit cell edges. In other words, there are $2$ effective minima enclosed by the cell, which would produce $2$ subbands. At present, this class has not been observed to form in the considered cases, but, in general,
we cannot rule out the possibility of this class forming at arbitrary moir\'e potentials. The model of this bandstructure is a simplified version of the one covered in Sec.~\ref{sec_tbm_b2}.
Here a superlattice similar to that of Fig.~\ref{fig_band2}(a2) is formed but with a single site in each non-empty superlattice cell. For this reason, there is a single type of state  in each cell: $\ket{A}$ in blue cells and $\ket{A}'$ in red cells. The Hamiltonian is
\begin{eqnarray}
\hat{H}
& = & \epsilon^\beta \sum_{i} (\hat{A}_i^\dagger \hat{A}_i + \hat{A}_i^{'\dagger} \hat{A}^{'}_i)
 \\
&& - W^{\beta}_A \sum_{\langle i,j \rangle \in R\leftrightarrow B} \hat{A}_i^\dagger \hat{A}^{'}_j
      \nonumber \\
&& - J^{\beta}_{A} \sum_{\langle i,j \rangle \in B\rightarrow B} \hat{A}_i^\dagger \hat{A}_j
      - J^{\beta}_{A} \sum_{\langle i,j \rangle \in R\rightarrow R} \hat{A}_i^{'\dagger} \hat{A}^{'}_j.
\nonumber
\end{eqnarray}
Note, here the on-site energy is directly $\epsilon^\beta$ and the hopping is not directional.
The latter can be diagonalized as $\hat{H}_A$ in Sec.~\ref{sec_tbm_b2}, and we find two bands
\begin{equation} \label{eq_classII_disp}
\begin{aligned}
\varepsilon^\beta_{a}(\mathbf{k}) = \epsilon^\beta - 2J^{\beta}_{A}(\cos k_x \ell + \cos k_y \ell ) \\ +4W^{\beta}_A\Big(\cos \frac{k_x \ell}{2} \cos \frac{k_y \ell}{2} \Big), \\
\varepsilon^\beta_{a'}(\mathbf{k}) = \epsilon^\beta - 2J^{\beta}_{A}(\cos k_x \ell + \cos k_y \ell ) \\ -4W^{\beta}_A\Big(\cos \frac{k_x \ell}{2} \cos \frac{k_y \ell}{2} \Big). \\
\end{aligned}
\end{equation}
These formulas are similar to Eqs.~(\ref{eq:band2a}) and (\ref{eq:band2ap}) for $J^{\beta}_{A;x}=J^{\beta}_{A;y}$ and thus $\varepsilon_{AA}^\beta(\mathbf{k})=\varepsilon_{A'A'}^\beta(\mathbf{k})$.

\subsection{Class III}
The next Class~III geometries in Fig.~\ref{fig_bandClasses} describe a wider variety of bands in twisted moir\'e potentials. For these cases, $4$ potential minima are now located within the unit cell. The minima are located outwith the green, superlattice cell, i.e. they are localised towards the unit cell centre. The resulting Class III bands can then be modelled using the procedures outlined in Sec.~\ref{sec_tbm_b345}.

\subsection{Class IV}
The final set of Class~IV bands in Fig.~\ref{fig_bandClasses} describes a more complex series of bands that must be described by a superlattice structure. There are again $4$ potential minima in the cell, but now located within the green, superlattice cells. In other words, inter-cell couplings dominate over intra-cell ones.
This situation can be treated using the superlattice scheme discussed in Sec.~\ref{sec_tbm_b2} for band $\beta=2$.


\section{Conclusions} \label{section_cnc}
In summary, we have discussed the single-particle spectrum of twisted moir\'e potentials across a range of potential depths, and described how effective tight-binding models are constructed in different bands.
If the moir\'e potentials are deep enough, each distinct minima of the potential contributes to unique sets of bands within the overall spectrum, with the exact form derived from the anisotropic harmonic approximation.
As the potential depth is decreased, couplings between potential minima become significant, thus lifting the harmonic degeneracies.
Localised Wannier states can still form at potential minima, allowing for bands and subbands to be characterised by $1$ of $4$ distinct classes of tight-binding geometries. The latter are determined by a hierarchy of energy scales encompassing strong local couplings, and weak couplings that support long-range coherence.
This allows us to build various effective tight-binding models, the parameters of which (tunnelling and onsite energies) can be fitted to the exact continuous-space spectrum.

The effective tight-binding models we derive allow us to understand the structure of energy bands, some of which break individually the four-fold rotation symmetry of the moir\'e potentials discussed here.
Our findings are also directly relevant to the many-body, bosonic counterpart of Hamiltonian~\eqref{eq_sph} with 2D contact interactions~\cite{johnstone2024weak}. In the strongly-interacting regime, spectral gaps can be mapped to insulating phases, and bands of states to compressible phases. For the compressible domains, we may have either a superfluid (SF) or normal fluid (NF), and the tunnellings determined serve as typical temperature scales where the system transitions from one to the other: If the temperature $T$ of the many-body system is such that $\kB T \lesssim J^\beta_\alpha$ for the corresponding band, then the compressible phase will be that of a SF. Otherwise, for $\kB T \gtrsim J^\beta_\alpha$, we will have a thermal, NF phase. Our results therefore illustrate the thermal stability of SF order in twisted moir\'e potentials. In realistic experiments with a typical $T\sim10nK$, we have $\kB T\sim10^{-2}\Er$ for most atomic species. For most of the low energy bands, we then have $J^\beta_\alpha\ll10^{-2}$, i.e.~SF order will generally not be stable for small chemical potentials $\mu/\Er$. There are, however, some exceptions, notably for moir\'e potentials with smaller unit cells, in which effective couplings are enhanced by virtue of the smaller separation between sites~\cite{johnstone2024weak}.

The framework and procedures that we have introduced to describe the bandstructure of twisted moir\'e potentials can also be applied to other, more general twisted potentials as well, e.g.~a superposition of $N$ periodic optical lattices with different twist angles between them. If this general potential has a periodic form, i.e.~moir\'e twisting angles, then different bands of states can be identified based off of the distribution of potential minima in the moir\'e cell. However, generally speaking, by increasing the number of superimposed lattices, we expect moir\'e lengths to increase, i.e.~tunnellings and SF order will be weaker. 
Finally, the physics of twisted moir\'e potentials can be linked to twisted bilayer or multilayer systems for strong interlayer couplings~\cite{meng2023atomic}.
This includes twisted bilayer graphene, which can be modelled using state-dependent, hexagonal rather than square optical potentials with a twist~\cite{cirac2019} and the procedures discussed here could be extended to such cases as well.

\begin{acknowledgments}
We thank Hepeng Yao and Shengjie Yu for fruitful discussions.
We acknowledge the CPHT computer team for valuable support.
This research was supported by
the Agence Nationale de la Recherche (ANR, project ANR-CMAQ-002 France~2030),
the program `Investissements d'Avenir'', the LabEx PALM (project ANR-10-LABX-0039-PALM), the IPParis Doctoral School and HPC/AI resources from GENCI-TGCC (Grant 2023-A0110510300).
\end{acknowledgments}

\appendix

\section{Geometrical Properties of Moir\'e Potentials} \label{app_geo}
Here, we provide further details and derivations on the geometrical properties of twisted moir\'e potentials. To begin, we discuss the origin of moir\'e angles. Given two square lattice vectors
\begin{equation} \label{eq_lt_V}
\begin{aligned}
\mathbf{a}_{m,n}/a & = m \hat{\mathbf{x}} + n \hat{\mathbf{y}}, \\ \mathbf{b}_{m',n'}/a & = m' R_\theta\hat{\mathbf{x}} + n'R_\theta \hat{\mathbf{y}},
\end{aligned}
\end{equation}
where $\hat{\mathbf{x}}$, $\hat{\mathbf{y}}$ are unit vectors along the $x$-, $y$-axis and $R_\theta$ is the rotation matrix with angle $\theta$
\begin{equation}
\begin{aligned}
R_\theta=
\begin{pmatrix}
\cos \theta & -\sin \theta \\
\sin \theta & \cos\theta
\end{pmatrix},
\end{aligned}
\end{equation}
a commensurate, or moir\'e angle is defined when the lattice vectors intersect, i.e.
\begin{equation}
\begin{aligned}
\mathbf{a}_{m,n} = & \, \mathbf{b}_{m',n'}, \\
\begin{pmatrix}
m \\
n
\end{pmatrix} = & \begin{pmatrix}
m' \cos\theta - n' \sin\theta \\
n'\cos\theta + m'\sin\theta
\end{pmatrix},
\end{aligned}
\end{equation}
from which it follows that
\begin{equation} \label{eq_lt_Vcs}
\begin{aligned}
\cos\theta = \frac{mm' + nn'}{m'^2 + n'^2}, \, \, \,\sin\theta = \frac{m'n - mn'}{m'^2 + n'^2}.
\end{aligned}
\end{equation}
The numbers in the numerators and denominators of Eq.~\eqref{eq_lt_Vcs}
\begin{equation} \label{eq_lt_Z}
\begin{aligned}
Z_1 & = mm' + nn',\\ Z_2 & = m'n - mn' \\ Z_3 & = m'^2 + n'^2,
\end{aligned}
\end{equation}
are all integers since $m$, $m'$ $n$ and $n'$ are integers.
The numbers $Z_1$, $Z_2$, and $Z_3$ thus form a Pythagorean triple, which imposes the constraint that $m$ and $n$ are coprime integers, with $m' = n$ and $n' = m$. Equation~\eqref{eq_lt_Vcs} can then be simplified as
\begin{equation}
\begin{aligned}
\cos\theta_{m,n} = \frac{2mn}{m^2 + n^2}, \, \, \,\sin\theta_{m,n} = \frac{n^2 - m^2}{m^2 + n^2},
\end{aligned}
\end{equation}
where $\theta_{m,n}$ is the moir\'e angle, as per Eq.~\eqref{eq_c_angle}.

Two twisted lattices will therefore intersect at the points $\mathbf{r}/a=(0,0)$ and $\mathbf{r}/a=(m,n)$, separated by a distance $\sqrt{m^2 + n^2}/a$. One would then expect, in general, that this distance is the period of the moir\'e lattice $\ell_{m,n}$. However, some care has to be taken when considering the parities of $m$ or $n$, and the uniqueness of $\theta_{m,n}$. The full set of moir\'e angles can be defined in the range $[0 \dots 45^\circ]$. Any moir\'e angle outside this range will have an equivalent angle to one within the range.
For example, $\theta_{2,1}\approx36.87^\circ$. We also have $\theta_{3,1}\approx53.13^\circ=-36.87^\circ$, i.e. $\cos\theta_{2,1} = \sin\theta_{3,1}$. Both $\theta_{2,1}$ and $\theta_{3,1}$ are geometrically equivalent, but care has to be taken in how the unit cell is defined with the different $m$ and $n$. Let us consider two cases: $m+n$ as an odd number and $m+n$ an even number. For $m+n$ odd, i.e. $\theta_{2,1}$ with $m+n=3$, $m^2 + n^2=5$ is also an odd and prime number. For $m+n$ even, i.e. $\theta_{3,1}$ with $m+n=4$, $m^2 + n^2=10$ is an even number. Clearly, the true moir\'e period must be $\ell_{2,1}=\sqrt{5}\equiv\ell_{3,1}=\sqrt{10/2}$. In summary, given a moir\'e angle $\theta_{m_o,n_o}$ where $m_o+n_o$ is an odd number, there is a geometrically equivalent moir\'e angle $\theta_{m_e,n_e}$ for $m_e+n_e$ as an even number, which are related via
\begin{equation}
\begin{aligned}
m_o^2 + n_o^2 = \frac{m_e^2 + n_e^2}{2}.
\end{aligned}
\end{equation}
Given this condition, the true moir\'e period $\ell_{m,n}$ is then written as 
\begin{equation}
\ell_{m,n}/a = \begin{cases}
\sqrt{(m^2 + n^2)/2},& \textrm{if } m+n\, \textrm{even},\\
\sqrt{m^2 + n^2},& \textrm{if } m+n \, \textrm{odd},
\end{cases}
\end{equation}
as per Eq.~\eqref{eq_m}.

By superimposing two square optical potentials, we can then generate a twisted moir\'e potential by tuning the twist angle to $\theta_{m,n}$. This generates an intricate superlattice structure, with distinct sets of potential minima in the moir\'e cell. The total number of distinct minima in a moir\'e potential is found to follow the relation given in Eq.~\eqref{eq_m}. Note, this relation has been modelled empirically. The formal derivation of the number of distinct minima requires solutions to the problem $\partial V (\mathbf{r}) / \partial x = \partial V (\mathbf{r}) / \partial y = 0$. However, to the best of our knowledge, no analytical solutions to this problem exist. Note, however, Eq.~\eqref{eq_m} always produces an integer number, which can be shown as follows.
As $m$ and $n$ are two coprime integers,
we may distinguish two cases:
(i)~one is an odd number and the other is an even number;
(ii)~both odd numbers.
For case~(i), $m$ and $n$ have different parities and we may suppose $m$ is odd and $n$ is even, without loss of generality.
Generally, for three integers $a,b,c$, it is easy to show that if $a\equiv b \mod{c}$, then $a^2\equiv b^2 \mod{c}$.
For $m$ as an odd number, $m\equiv 1 \mod{4}$ or $m\equiv 3 \mod{4}$. Since $3^2\equiv 1 \mod{4}$, we always have $m^2\equiv 1 \mod{4}$.
For an even number $n$, $n^2\equiv 0 \mod{4}$. Since $m+n$ is odd, according to Eq.~\eqref{eq_l_length}, $\ell_{m,n}/a=\sqrt{m^2+n^2}$. So $\ell_{m,n}^2/a^2=m^2+n^2\equiv 1 \mod{4}$,
which means that $\mathcal{M}_{m,n}$ [Eq.~\eqref{eq_m}] is an integer. For case (ii), $m$ and $n$ are both odd numbers. For $m$ as an odd number, $m\equiv 1\text{ or } 3\text{ or } 5\text{ or } 7 \mod{8}$, then $m^2\equiv1^2\equiv 3^2\equiv 5^2\equiv 7^2\equiv 1 \mod{8}$. Similarly, for $n$ as an odd number, we also have $n^2\equiv 1 \mod{8}$. Then, we have $m^2+n^2\equiv 2 \mod{8}$ and thus $\frac{m^2+n^2}{2}\equiv 1 \mod{4}$. As $m+n$ is even, according to Eq.~\eqref{eq_l_length}, $\ell_{m,n}^2/a^2=\frac{m^2+n^2}{2}$. So in this case also, $\mathcal{M}_{m,n}$ [Eq.~\eqref{eq_m}] must also be an integer.

\section{Eigenstate Basis to Lattice Site Basis}\label{app:OppositeJs}
When constructing tight-binding models, it is also of interest to consider their representation in the basis of lattice sites, rather than the eigenstate basis. These representations are also important for understanding the origin of anisotropic tunnelling rates in the eigenstate basis.
For band $\beta=1$ (more generally Classes I and II bands in Fig.~\ref{fig_bandClasses}), each moir\'e cell or superlattice cell will respectively contain $1$ lattice site, hence no conversion is necessary for these bands. However, for the other bands (Classes III and IV), the situation is different, since the intra-cell Hamiltonians contain more than $1$ unique eigenstate.

\subsection{Class III Bands} \label{app_latticeIII}
\begin{figure}[t!]
	\centering
	\makebox[0pt]{\includegraphics[width=0.7\linewidth]{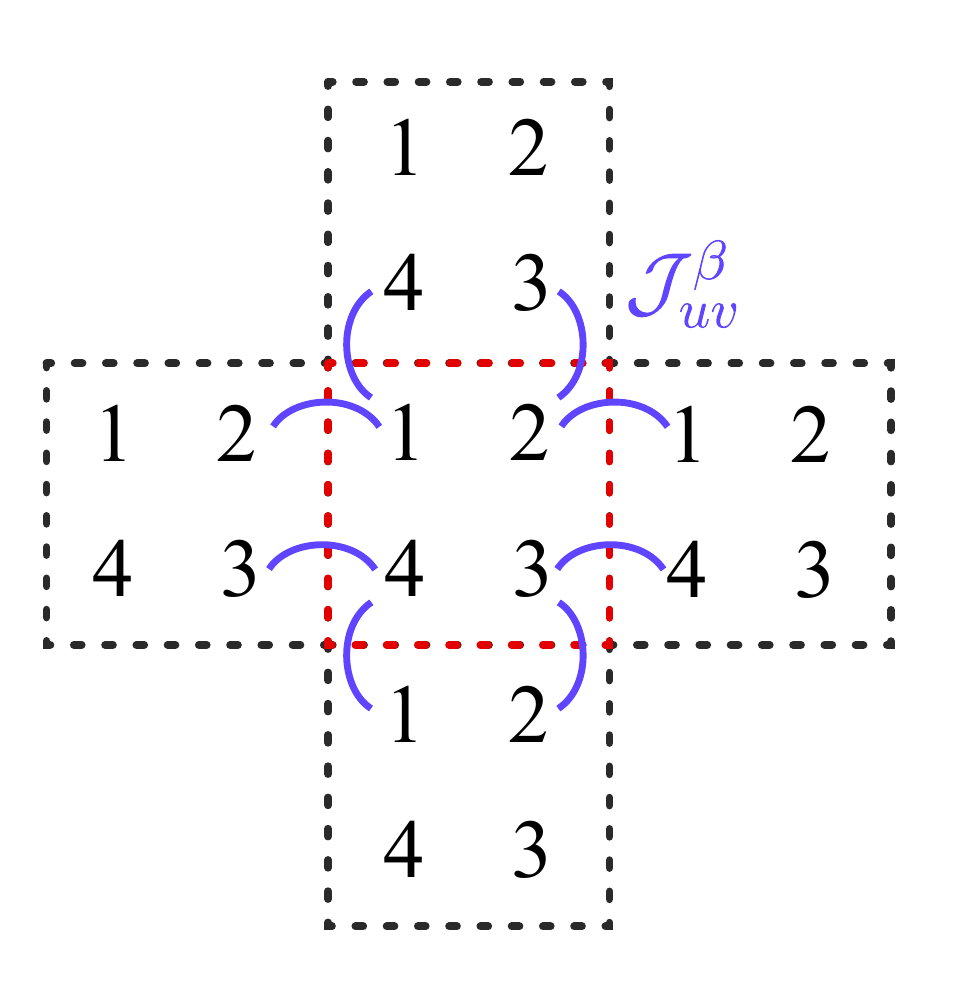}}
	\caption{Origin of inhomogeneous eigenstate couplings. Here, we show sites in the lattice site basis (black numbers), in a red central cell and $4$ black neighbouring cells. Real-space tunnellings $\mathcal{J}^\beta_{u,v}$ between a site $u$ in the red cell and $v$ in a black cell must preserve the $4$-fold rotational symmetry of the moir\'e potential. As an example, this implies that all of the marked, blue tunnellings must be equal. These conditions then impose constraints on the corresponding eigenstate-basis couplings $J_\alpha^\beta$, according to Eq.~\eqref{eq_tunnel1}.}
	\label{fig_band3_eig}
\end{figure}

As discussed in Sec.~\ref{sec_tbm_b345}, tunnelling rates $J_{B;i,j}^{\beta'}$ must be equal for a Class III band, but sign flipped to preserve global $4$-fold rotational symmetry. To show this, we rewrite Hamiltonian~\eqref{eq_h_lt_full} in the basis of lattice sites, rather than the eigenstate basis. This can be done by representing the eigenstates in Eq.~\eqref{eq_eigVec} with operators that create/destroy particles at one of the $4$ intra-cell sites. For example, we have $\hat{a}_i=(\hat{s}_i+\hat{t}_i+\hat{u}_i+\hat{v}_i)/2$, where the operators $\hat{s}_i$, $\hat{t}_i$, $\hat{u}_i$, and $\hat{v}_i$ destroy a particle at sites $1$, $2$, $3$, and $4$ from Fig.~\ref{fig_band3_eig}, respectively. The other eigenstate operators $\hat{B}_i$, $\hat{C}_i$, $\hat{d}_i$ can be represented in a similar way. When we convert Hamiltonian~\eqref{eq_h_lt_full} to the lattice site basis, we will then have terms such as $\hat{s}^\dagger_i \hat{t}_j$, e.g.~a tunnelling operator from site $1$ to site $2$ of an adjacent cell, etc. Each site of a cell couples to $4$ sites of an adjacent cell, producing $16$ distinct tunnelling rates and onsite terms in the lattice site basis. For the tunnelling operators, the coefficients between a site $u$ in one cell and a site $v$ in an adjacent cell can be computed as
\begin{equation} \label{eq_tunnel1}
\begin{aligned}
\mathcal{J}^\beta_{u,v} = J^\beta_a a_u a_v + J^\beta_d d_u d_v + J^\beta_B (B_u B_v + C_u C_v) \\ + \, W^{\beta}_{B;u,v} (B_u C_v + C_u B_v)
\end{aligned}
\end{equation}
where the coefficients $a_{i}$ are the elements of eigenstate $\ket{a}$ at index $i$. Likewise, the intra-cell/onsite couplings between sites $u$ and $v$ will be
\begin{equation}
\begin{aligned}
\eta^\beta_{u,v} = E^\beta_a a_u a_v + E^\beta_d d_u d_v + E^\beta_B (B_u B_v + C_u C_v).
\end{aligned}
\end{equation}
One can then easily verify that for each $u$ and $v$, $\eta^\beta_{u,v}$ will simply be equal to one of the intra-cell parameters of the original, decoupled Hamiltonian in Eq.~\eqref{eq_h_lt}, i.e. $\eta^\beta_{1,1}=\epsilon^\beta$, $\eta^\beta_{1,2}=\mathcal{I}^\beta$, etc. Based on the geometry of the moir\'e potential, the real-space tunnellings $\mathcal{J}^\beta_{u,v}$ must preserve $4$-fold rotational symmetry, see Fig.~\ref{fig_band3_eig}. As an example, this implies that $\mathcal{J}^\beta_{1,4}$ across the $+y$ direction must be equal to $\mathcal{J}^\beta_{3,4}$ across the $+x$ direction. Each tunnelling can be explicitly written as
\begin{equation}
\begin{aligned}
\mathcal{J}^\beta_{1,4} = \frac{1}{2}J^\beta_a - \frac{1}{2}J^\beta_d - \frac{1}{4}W^{\beta}_{B;1,4}, \\ \mathcal{J}^\beta_{3,4} = \frac{1}{2}J^\beta_a - \frac{1}{2}J^\beta_d + \frac{1}{4}W^{\beta}_{B;3,4}.
\end{aligned}
\end{equation}
For rotational symmetry to be preserved, we must have $W^{\beta}_{B;1,4}=-W^{\beta}_{B;3,4}$. In other words $W^{\beta}_{B;x}=-W^{\beta}_{B;y}=W^{\beta}_{B}$, where $x,y$ denotes the $x$ and $y$ directions.

\subsection{Class IV Bands} \label{app_latticeIV}
A similar process can also be performed for Class IV bands, such as $\beta=2$.
Each superlattice cell has two sites, giving $4$ unique couplings between two different cells. The real-space tunnelling rates between the different types of cells are given by
\begin{equation}
\begin{aligned}
\mathcal{J}^{B\rightarrow B}_{u,v} &= J^\beta_{A;u,v} A_u A_v + J^\beta_{B;u,v} B_u B_v \\
\mathcal{J}^{R\rightarrow R}_{u,v} &= J^\beta_{A';u,v} A_u A_v + J^\beta_{B';u,v} B_u B_v \\
\mathcal{J}^{B\leftrightarrow R}_{u,v} &= W^{\beta}_{A} A_u A_v + W^{\beta}_{B} B_u B_v
\end{aligned}
\end{equation}
The intra-cell/onsite couplings can also be expressed as
\begin{equation}
\begin{aligned}
\eta^\beta_{u,v} = 2(E^\beta_A A_u A_v + E^\beta_B B_u B_v),
\end{aligned}
\end{equation}
which are again equal to the intra-cell parameters of Hamiltonian~\eqref{eq_h_lt2}. The couplings $\mathcal{J}^{B\rightarrow B}_{u,v}$ and $\mathcal{J}^{R\rightarrow R}_{u,v}$ have a similar form and corresponding unit cell, but are aligned across different directions. In order to preserve $4$-fold rotational symmetry, we must also then have $J^{\beta}_{A;x}=J^{\beta}_{A';y}$, $J^{\beta}_{A;y}=J^{\beta}_{A';x}$, $J^{\beta}_{B;x}=J^{\beta}_{B';y}$ and $J^{\beta}_{B;y}=J^{\beta}_{B';x}$.

\section{Extracting Tight-Binding Parameters From Continuous Dispersions}\label{app:tbm_val}
Here, we outline the extraction of tight-binding parameters for the different classes of bands in Fig.~\ref{fig_bandClasses}.

\subsection{Class I}
Class I bands, such as band $\beta=1$ in Sec.~\ref{sec_tbm_b1} for $\theta_{3,5}$, have a standard cosine dispersion as per Eq.~\eqref{eq_k_sq}. The average and amplitude of the dispersion relations can be calculated to extract $\epsilon^\beta$ and $J^\beta$.

\subsection{Class II}
For Class II bands in Eq.~\eqref{eq_classII_disp}, we have $2$ equations and $3$ parameters to fit. First, we consider the average dispersion
\begin{equation}
\begin{aligned}
\frac{\varepsilon^\beta_{a}(\mathbf{k})+\varepsilon^\beta_{a'}(\mathbf{k})}{2} = \epsilon^\beta - 2J^{\beta}_{A}(\cos k_x \ell + \cos k_y \ell ),
\end{aligned}
\end{equation}
from which $\varepsilon^\beta_{a}$ and $J^{\beta}_{A}$ can be found as per a Class I band, independent from $W^{\beta}_{A}$.
Instead, $W^{\beta}_{A}$ is calculated by fitting the difference of dispersions,
\begin{equation}
\begin{aligned}
\frac{\varepsilon^\beta_{a}(\mathbf{k})-\varepsilon^\beta_{a'}(\mathbf{k})}{2} = -4W^{\beta}_A\Big(\cos \frac{k_x \ell}{2} \cos \frac{k_y \ell}{2} \Big),
\end{aligned}
\end{equation}
which is independent of $\varepsilon^\beta_{a}$ and $J^{\beta}_{A}$.

\subsection{Class III} \label{app:tbm_val_c3}
For Class III bands, such as those considered in Sec.~\ref{sec_tbm_b345}, we must extract $3$ intra-cell parameters and $4$ inter-cell couplings from $4$ dispersions. For the $2$ non-degenerate subbands $a$ and $d$, eigenvalues $E_{a,d}^\beta$ and tunnellings $J_{a,d}^\beta$ can extracted from the standard cosine dispersions as per class~I. For degenerate bands $b$ and $c$, we consider average and difference functions
\begin{equation} \label{eq_disp_avg}
\begin{aligned}
\frac{{\varepsilon}^\beta_{b}(\mathbf{k})+{\varepsilon}^\beta_{c}(\mathbf{k})}{2} & =  E_B^\beta - 2J^\beta_B(\cos k_x \ell + \cos k_y \ell) = \varepsilon^\beta_B(\mathbf{k}),
\end{aligned}
\end{equation}
and
\begin{equation}
\begin{aligned}
\frac{{\varepsilon}^\beta_{b}(\mathbf{k})-{\varepsilon}^\beta_{c}(\mathbf{k})}{2} & = - 2W^{\beta}_{B}(\cos k_x \ell - \cos k_y \ell) = \varepsilon^{\beta'}_B(\mathbf{k}),
\end{aligned}
\end{equation}
which yield $E_B^\beta$ and $J^\beta_B$, on the one hand, and $W^{\beta}_{B}$, on the other hand, from independent fits.
The effective tunnelling energies $J_b^\beta$ and $J_c^\beta$ associated to the final dispersion relations are then found from Eq.~(\ref{eq:Js}).
Finally, the eigenvalues $E^\beta_\alpha$ of the intra-cell Hamiltonian are simply the energy shifts to each dispersion relation, allowing for the intra-cell parameters to be calculated as
\begin{equation}
\begin{aligned}
\epsilon^\beta & = \frac{1}{4}(E^\beta_a + E^\beta_d + 2E^\beta_B), \\
\mathcal{I}^\beta & = \frac{1}{4}(E^\beta_a - E^\beta_d), \\
\mathcal{I}^{\beta'} & = \frac{1}{4}(-E^\beta_a - E^\beta_d + 2E^\beta_B), \\
\end{aligned}
\end{equation}
which is equivalent to Eq.~(\ref{eq:EigenVal}).

\subsection{Class IV} \label{app:tbm_val_c4}
Finally,
for Class IV bands, such as the one considered in Sec.~\ref{sec_tbm_b2},
we first consider the average dispersion for each degenerate set of subbands,
\begin{equation} \label{eq_disp_AB2_avg}
\begin{aligned}
\frac{{\varepsilon}^\beta_{a}(\mathbf{k}) + {\varepsilon}^\beta_{a^{'}}(\mathbf{k})}{2} = \frac{{\varepsilon}^\beta_{AA}(\mathbf{k}) + {\varepsilon}^\beta_{A^{'}A^{'}}(\mathbf{k})}{2} = \\
E^\beta_{A} - 2(J^{\beta}_{A;x}+J^{\beta}_{A;y})(\cos k_x \ell + \cos k_y \ell ).
\end{aligned}
\end{equation}
Here we focus on subbands $A-A'$, and similar formulas are found for subbands $B-B'$.
In this way, we can fit the values for $E^\beta_A$ and $E^\beta_B$,
and determine the intra-cell parameters from the energy shifts/eigenvalues
\begin{equation}
\begin{aligned}
\epsilon^\beta & = \frac{1}{2}(E^\beta_A + E^\beta_B), \\
\mathcal{I}^\beta & = \frac{1}{2}(E^\beta_A - E^\beta_B),
\end{aligned}
\end{equation}
as well as the widths $J^{\beta}_{A;x}+J^{\beta}_{A;y}$ and $J^{\beta}_{B;x}+J^{\beta}_{B;y}$.
Next, we take the differences,
\begin{equation} \label{eq_disp_AB2_diff}
\begin{aligned}
{\varepsilon}^\beta_{a}(\mathbf{k}) - {\varepsilon}^\beta_{a^{'}}(\mathbf{k}) = \sqrt{\left(\varepsilon^\beta_{AA}(\mathbf{k})-\varepsilon^\beta_{A'A'}(\mathbf{k})\right)^2 + 4\varepsilon^{\beta}_{AA'}(\mathbf{k})^2 }
\end{aligned}
\end{equation}
First, we consider a $\mathbf{k}$-direction in which $\varepsilon^{\beta}_{AA'}(\mathbf{k})=0$,
e.g.~$k_{y}/\ell=\pi$. We then have
\begin{equation}
\begin{aligned}
\tilde{\varepsilon}^\beta_{a}(\mathbf{k}) - \tilde{\varepsilon}^\beta_{a^{'}}(\mathbf{k}) =
\left\vert\varepsilon^\beta_{AA}(\mathbf{k})-\varepsilon^\beta_{A'A'}(\mathbf{k}) \right\vert \\
= -2\left\vert (J^{\beta}_{A;x}-J^{\beta}_{A;y}) ( \cos k_x \ell + \cos k_y \ell ) \right\vert,
\end{aligned}
\end{equation}
from which we may extract the widths $J^{\beta}_{A;x}-J^{\beta}_{A;y}$.
We may then solve for the individual couplings $J^{\beta}_{A;x}$
and $J^{\beta}_{A;y}$
by using the results from Eq.~\eqref{eq_disp_AB2_avg}.
Finally, we can consider a $\mathbf{k}$-direction in which $\varepsilon^\beta_{AA}(\mathbf{k})=\varepsilon^\beta_{A'A'}(\mathbf{k})$,
i.e.~$k_x = k_y$. This yields
\begin{equation}
\begin{aligned}
{\varepsilon}^\beta_{a}(\mathbf{k}) - {\varepsilon}^\beta_{a^{'}}(\mathbf{k}) = 2\varepsilon^\beta_{AA'}(\mathbf{k}) \\
= -8W^{\beta}_A\Big(\cos \frac{k_x \ell}{2} \cos \frac{k_y \ell}{2} \Big),
\end{aligned}
\end{equation}
allowing for $W^{\beta}_A$
to be extracted from the widths.
The parameters for the $B-B'$ subbands are found similarly.


%

\end{document}